\documentclass[a4paper,fleqn]{cas-sc}

\usepackage[numbers,sort&compress]{natbib}
\usepackage{graphicx}
\usepackage{float}
\usepackage{amsmath}
\usepackage{comment}
\usepackage{xcolor}
\usepackage{url}
\usepackage{subcaption}
\usepackage{ragged2e}

\def\tsc#1{\csdef{#1}{\textsc{\lowercase{#1}}\xspace}}
\tsc{WGM}
\tsc{QE}

\begin{document}
\let\WriteBookmarks\relax
\def\floatpagepagefraction{1}
\def\textpagefraction{.001}

\shorttitle{}    

\shortauthors{Preeti Sar, Sebastian De Pascuale, Harry G. Dudding, Gary M. Staebler}  

\title [mode = title]{A machine learning framework for developing quasilinear saturation rules for tokamak turbulent transport}  



%

\author[1]{Preeti Sar}[type = editor, orcid = 0000-0002-4937-7023]
\cormark[1]
\ead{sarp@ornl.gov}
\credit{}
\affiliation[1]{organization={Fusion Energy Division, Oak Ridge National Laboratory},
            city={Oak Ridge},
            postcode={37830}, 
            state={TN},
            country={USA}}
\cortext[1]{Corresponding author}

\author[1]{Sebastian De~Pascuale}[orcid = 0000-0001-7142-0246]
\ead{depascuales@ornl.gov}
\credit{}

\author[2]{Harry G. Dudding}[orcid = 0000-0002-3413-0861]
\ead{harry.dudding@ukaea.uk}
\credit{}
\affiliation[3]{organization={United Kingdom Atomic Energy Authority},
            city={Abingdon},
            postcode={OX14 3DB}, 
            state={Oxfordshire},
            country={UK}}            

\author[1]{Gary M. Staebler}[orcid = 0000-0002-1944-1733]
\ead{staeblergm@ornl.gov}
\credit{}


\begin{abstract}
A new neural network model for a quasilinear saturation rule has been developed to map linear gyrokinetic data to nonlinear saturated potential magnitudes in order to predict the total energy and particle fluxes expected from nonlinear gyrokinetic simulations. The training dataset is taken from the high resolution simulation database of linear and nonlinear simulations generated with the CGYRO code for developing the SAT3 model. This new model, named NN-SAT3, overall is able to capture the 1D saturated potential magnitudes of the dataset more accurately than SAT3, as depicted by lower percentage errors in the peak values of the 1D saturated potentials, as well as the locations of the peaks in binormal wavenumber $k_y$ space. The resulting fluxes also had smaller deviations from the nonlinear CGYRO data as compared to previous saturation models such as SAT0 - SAT3. Consistent with SAT3, NN-SAT3 is able to recreate the anti-gyroBohm scaling of fluxes for the considered TEM-dominated cases as well as better predictions near the point of critical threshold
when plotted against temperature gradients.
\end{abstract}




\maketitle



\justifying

\vspace{0.5cm}
This manuscript has been authored in part by UT-Battelle, LLC, under contract DE-AC05-00OR22725 with the US Department of Energy (DOE). The publisher acknowledges the US government license to provide public access under the DOE Public Access Plan. \small(\url{ http://energy.gov/downloads/doe-public-access-plan})

\section{Introduction}
Although the tokamak design provides a favorable basis for  magnetic confinement fusion, it has not yet demonstrated sufficiently sustained confinement at plasma parameters necessary for economical energy generation. Turbulent fluctuations within the plasma are typically the dominant cause of the transport of energy and particles observed in a tokamak, driving loss of plasma confinement seen in experiments. The steep gradients in density and temperature of the plasma influence the growth of perturbative oscillations, which interact nonlinearly and reach saturation with a steady level of fluctuation. Optimization of the core performance requires accurate modeling paradigms, which can help us understand how to achieve improved performance and control in real time. 

State of the art turbulent transport models are based on nonlinear gyrokinetic theory, which is the  most accurate method to model plasma turbulence, its accuracy having been tested by various experiments \cite{rhodes2011mode, holland2011advances, told2013characterizing, citrin2014ion, bonanomi2015trapped, creely2017validation, freethy2018validation, citrin2022integrated}. But this accuracy comes at the cost of computational expense, with a single local simulation taking at least on the order of $10^4-10^5$ CPUh \cite{bourdelle2015core}. This large computational cost makes nonlinear gyrokinetics unsuitable for integrated modeling, and hence alternative approaches to turbulent flux transport calculations have been proposed. This computational challenge has led to the development of reduced models.

Reduced transport models aim to predict with sufficient fidelity the turbulent transport fluxes obtained from first-principles gyrokinetic simulations. A class of such reduced models called quasilinear models, such as Trapped Gyro Landau Fluid (TGLF) \cite{staebler2005gyro, staebler2007theory,kinsey2008first} and QuaLiKiz \cite{bourdelle2015core}, are  used for faster simulation times. The TGLF model predicts the energy, particle and momentum fluxes of gyrokinetic simulations on the order of $10$ CPU-seconds per radial data point, enabling faster predictions. These quasilinear models solve for the linear response of the plasma instabilities, instead of the expensive calculations of fluxes in the nonlinear gyrokinetic system. These linear quantities are then combined with an estimation of the magnitude of the saturated nonlinear potentials to provide a calculation of the fluxes in much less time. In TGLF, the linear eigenvalues and eigenmodes of a system of gyro-fluid equations \cite{staebler2005gyro} are used to evaluate the quasilinear weights of the transport fluxes. The methods used for estimating the saturated potential magnitude are called saturation rules. Some examples of saturation rules that have been developed using ad hoc and semi-analytical methods are, SAT0 \cite{kinsey2008first}, SAT1 \cite{staebler2016role}, SAT2 \cite{staebler2021verification}, and SAT3 \cite{dudding2022new}. These saturation rules are tuned to specific nonlinear simulation datasets and combined with the linear physics of a quasilinear model such as TGLF to produce reduced models such as TGLF-SAT1, TGLF-SAT2. Both the TGLF and QuaLiKiz models have been extensively validated against nonlinear gyrokinetic codes thus verifying their accuracy for integrated modeling \cite{staebler2021verification, casati2009validating}. In addition, the linear eigenmodes of the TGLF model have been validated with a large database of gyrokinetic linear stability calculations \cite{staebler2005gyro}.

Control applications running in real-time require models that are capable of providing accurate solutions within milliseconds, and so even reduced quasilinear models are considered slow given that whole device transport simulations typically require thousands of evaluations. Hence machine learning surrogate models have been developed to calculate the total flux by using training datasets generated from quasilinear models, which given the faster simulation time over nonlinear gyrokinetic codes, aids in generating large datasets for developing accurate models. For example TGLF-NN \cite{neiser2022database, meneghini2017self} trains neural networks to predict the output flux of TGLF, and QuaLiKiz-NN \cite{citrin2015real,van2020fast} is trained to predict the QuaLiKiz output flux with specific saturation rules. Models like TGLF-SINN \cite{cao2025tglf} can train neural networks with a much smaller dataset than needed by TGLF-NN, and also without needing to clip negative fluxes to zero, unlike the QuaLiKiz-NN. There is also work done to build machine learning models to directly determine the eigenvalues of quasilinear models to better capture dominant modes due to the stiffness of turbulent transport, which can then be used to develop saturation models to calculate the total flux. For example, in Fransson et al \cite{fransson2023fast}, a neural network surrogate has been developed for the eigenvalues of the QuaLiKiz model. 

 By making the saturation model a function of the linear quantities, the high degrees of freedom in the input parameters of the linear stability calculation, which are complex scans over many plasma parameters, are reduced to a lower dimensional input space. This ensures that even though the saturation model is fit to a small number of nonlinear gyrokinetic runs, it is applicable to a wide range of input parameters interpolated within the linear input space which could be outside of the training set. Developing a neural network model of the saturation rule is motivated by the discrepancy in hand-fit empirical models with limited gyrokinetic simulations across a range of plasma parameters, which suggests room for improvement using a machine learning approach. Hence, our method depends only on linear quantities from linear gyrokinetic data to build a neural network model for developing a saturation rule, which then maps the turbulent fluxes more accurately than previous saturation models and also recreates certain flux characteristics such as anti-gyroBohm scaling, where the fluxes of certain TEM-dominated cases scale inversely with isotope mass \cite{belli2019reversal, garcia2022modelling}. This characteristic was shown in \cite{candy2016high} using the CGYRO gyrokinetic code, such that when the equilibrium density gradient $a/L_n$ was increased from a baseline \cite{waltz1997gyro} regime dominated by ion temperature gradient (ITG) turbulence to one dominated by trapped electron mode (TEM) turbulence, the flux scaling was reversed.

The aim of this work is to develop a new machine learning framework for developing quasilinear saturation rules of turbulent transport from linear and nonlinear gyrokinetic simulation data. This paper is organized as follows: section 2 contains a discussion on turbulent plasma fluxes, the structure of quasilinear models, as well as the detail of the gyrokinetic database used in this work, which was created for developing the SAT3 model. Section 3 describes the development of the new neural network saturation rule NN-SAT3, which is trained on the SAT3 simulation database. In section 4, the results of the NN-SAT3 saturation rule are presented and analyzed against the nonlinear (NL) CGYRO data and the SAT3 model. Section 5 tests the accuracy of the NN-SAT3 and SAT3 models outside of their training database by using linear input quantities generated by the Gyro Fluid Solver (GFS) \cite{staebler2023flexible}. Finally, the conclusion section will summarize the work done as well as briefly mention some avenues for future research.

\section{Turbulent Fluxes and Simulation Database}
 In this work, reduced models will be referred as <linear inputs>–<saturation rule>, to specify the constituents of each model. The linear eigenmodes computed with CGYRO are combined with a saturation rule to calculate the total turbulent fluxes. Thus CGYRO-SAT3 refers to SAT3 with the reduced linear inputs of TGLF and similarly, the NN-SAT3 saturation rule with the linear inputs of GFS is GFS-NN-SAT3. The expression for the total flux is written as a 1D sum of certain terms over the binormal wavenumber $k_y$. The final form for the structure of fluxes $Q_s$ (for species $s$) in quasilinear models, as shown in \cite{dudding2022new}, is given by
\begin{align}
Q_s = 2\sum_{k_y>0}\Lambda_{s,k_y}W^L_{s,k_y}F_{s,k_y}\left[\dfrac{\left\langle|\delta \hat{\phi}_{k_y}|^2\right\rangle_{x,\theta,t}}{\Delta k_y}\right]\Delta k_y,
\label{ql3}
\end{align}
which is comprised of five parts. $\Lambda_{s,k_y} = W^{NL}_{s,k_y}/W^L_{s,k_y}$ is the quasilinear approximation (QLA) function, which is a ratio of the nonlinear weights to the linear weights. The nonlinear weight $W^{NL}_{s,k_y}$ represents the average phase relation between the potential and pressure fluctuations during the nonlinear phase of the turbulence. The linear weight $W^L_{s,k_y}$ is similarly defined for the fluctuations in the linear gyrokinetic system. The 1D saturated potentials $\langle|\delta\hat{\phi}_{k_y}|^2\rangle_{x,\theta,t}$ are the squared Fourier amplitudes of the potential averaged over the radial direction $x$, parallel-to-field coordinate $\theta$ and time $t$. The binormal grid spacing, $\Delta k_y$, sets the discretization of the simulations. In reduced modeling, the potentials $\langle|\delta \hat{\phi}_{k_y}|^2\rangle_{x,\theta,t}$ are normalized to the binormal grid spacing $\Delta k_y$, which is shown to be invariant under a change of grid resolution once past the point of convergence in nonlinear gyrokinetic simulations \cite{dudding2022new}. $F_{s,k_y}$ describes the effect of $E\times B$ shear, whose value in this work is equal to $1.0$, as no finite $E\times B$ shear cases are included in the training database. Going forward, $\langle|\delta \hat{\phi}_{k_y}|^2\rangle_{x,\theta,t}$ will be denoted as $\phi_{k_y}^2$ for brevity in notation.

For training the neural network model, the simulation database prepared for creating the SAT3 model \cite{dudding2022new} is used. Hence, the new neural network model is called the NN-SAT3 saturation model. It must be made clear that this model is not the neural network surrogate of SAT3, but instead is a neural network fit to the same nonlinear turbulence database as the SAT3 model, improving on its fidelity. As part of the work for developing SAT3, a database of 43 nonlinear gyrokinetic simulations over scans of plasma parameters centered on the GA-standard (GA-std) case \cite{candy2009unified} and their corresponding linear simulations was generated using CGYRO \cite{candy2016high}. 

The database of parameter scans is primarily centered around the baseline GA-std case, which is defined on circular Miller flux surface geometry \cite{kinsey2010trapped} with parameters of $a/L_{T_i}$ = $a/L_{T_e} = 3.0$, $a/L_n = 1.0$, $T_i/T_e = 1.0$, $\hat{s} = 1.0$, $q = 2.0$, $(a/c_s)\nu_{ee} = 0.1$ and $r/R_0 = 1/6$. Here, $a/L_n$ is density gradient, where $a$ is the tokamak minor radius for the last closed surface, $a/L_{T_i} = a/L_{T_e}$ is temperature gradient where changes were kept the same between the ions and electrons, $\hat{s}$ is magnetic shear, $(a/c_s)\nu_{ee}$ is collisionality, $\kappa$ is elongation, $q$ is safety factor, $\Delta = dR_0/dr$ is Shafranov shift, $r/R_0$ is ratio of the tokamak minor radius of a given closed surface to the major radius, and $T_i/T_e$ is ratio of ion and electron temperature. The exact values scanned can be found in table 1 of \cite{dudding2022new}. The three isotopes that have been simulated are Hydrogen (H), Deuterium (D) and Tritium (T), with $m_i/m_D$ values of 0.5, 1.0 and 1.5 respectively, for a normalized gyroradius of $\rho_i = \sqrt{m_i/m_D}\rho_{unit}$ for each isotope. Here $m_i$ is the ion mass and $m_D$ is the deuterium mass. The deuterium sound speed is $c_s = \sqrt{T_e/m_D}$ and the reference gyroradius $\rho_{unit}$ is given by $\rho_{unit} = \sqrt{m_DT_e}/eB_{unit}$. $B_{unit}$ is an effective magnetic field given by
\begin{align}
    B_{unit} = \frac{q(r)}{r}\frac{d\psi}{dr},
\end{align}
where $\psi$ is the poloidal flux divided by $2\pi$, $q$ is the safety factor and $r$ denotes a given flux-surface. The database used in this work is available via an open repository \cite{citekey}.

The core of the SAT3 saturation rule describes an exponential relationship between the 1D saturated potential and the radial width of the spectrum. In the SAT3 model, the peak value of the saturated potential is determined by the saturation level, which is mapped as a linear regression fit  to combinations of linear physics quantities such as $\gamma_{max}$ and $k_{max}$, which are representative quantities of the growth rate spectrum. The width of the saturated potential is determined by the spectral shape which follows a quadratic form, the coefficients of which are fitted linearly to $k_{max}$. Here, $k_{max}$ is defined with respect to the linear gyrokinetic simulations as
\begin{equation}
    \frac{d}{dk_y}\left(\frac{\gamma_{k_y}}{k_y}\right)\Bigg |_{k_y = k_{max}} = 0. 
\end{equation}
The corresponding growth rate $\gamma$ evaluated at $k_{max}$ gives $\gamma_{max}$. For a full discussion, see \cite{dudding2022new}. 

\section{Machine Learning Methodology}
The regression method of SAT3 is relaxed via a neural network framework to identify a combination of free parameters that would accurately map the linear quantities to the nonlinear saturated potentials. In this work, the entire binormal wavenumber spectra of the 1D saturated potential is considered, represented by a nonlinear multi-layer perceptron neural network fit. We name this new model the NN-SAT3 model, given that this is a neural network model trained on the SAT3 database. This work was carried out in Python $3.12.10$, and the neural network was implemented using the Pytorch library. 

\subsection{Architecture} \label{arch}
\begin{figure}[!ht]
    \centering
        \includegraphics[width=1.0\linewidth]{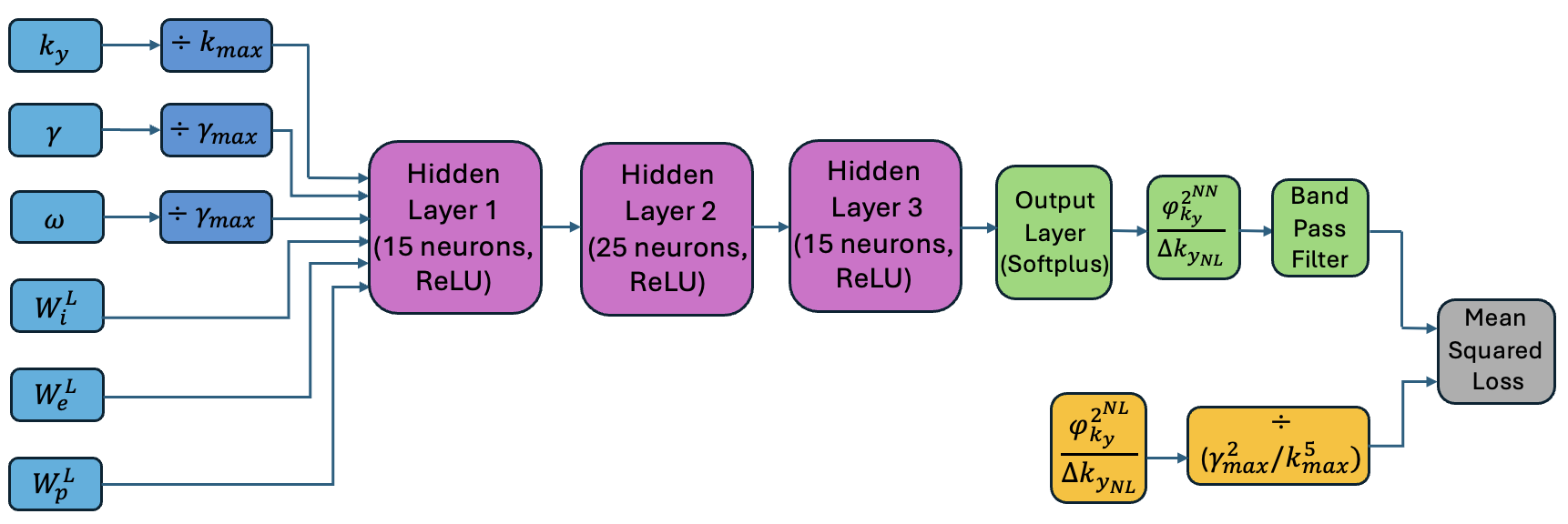}
        \caption{Neural network architecture mapping linear inputs to nonlinear 1D saturated potential. Blue blocks are the input layer along with the normalizations, pink blocks are the hidden layers, green blocks are the output layer along with the output modifications, yellow blocks are the target 1D saturated potential along with its normalization, and grey block is the mean squared loss function.}
    \label{nn}
\end{figure}

A multi-layer perceptron (MLP) with six inputs, three hidden layers and one output is used as shown in Figure \ref{nn}. The first hidden layer has $15$ neurons, the second layer has $25$ neurons and the third layer has $15$ neurons. The connections across every layer has associated weights, and every hidden and output layer uses a bias term, giving 
\begin{align}
   \text{Number of free parameters} = \sum_{i = 1}^{L} (n_{i-1} +1)n_i,
\end{align}
where layer $0$ is the input layer, layers 1 to $L-1$ are hidden layers, layer $L$ is the output layer, $n_i$ is the number of neurons in current layer $i$ and $n_{i-1}$ is the number of neurons in previous layer $i-1$. With 6 inputs, 3 hidden layers with 15, 25 and 15 neurons in each, and 1 output, this amounts to a total of $911$ degrees of freedom. The six inputs are the binormal wavenumber $k_y$, growth rate $\gamma$, frequency $\omega$, and linear weights for ion energy $W^L_i$, electron energy $W^L_e$ and particle $W^L_p$. For each of the 43 cases, there are 40 $k_y$ per CGYRO run giving a total length of $1720$ $k_y$ discretizations. The first three indices of $k_y$ are not used in training the neural network. This is because the 1D saturated potentials increase towards the zonal potential at $k_y = 0$ (zonal mode), whereas the linear weights are very small for low $k_y$ and hence there is not much flux contribution in those areas, removing the need to fit the uptick in the 1D saturated potentials for low $k_y$. Hence, the 37 $k_y$ indices for each input leads to a total length of $1591$ for all the $43$ cases and so a total of 6 inputs gives $9546$ input parameters. Therefore $911$ free parameters in the neural network are fitted to $9546$ input parameters, ensuring that the model is not over-fitted. ReLU activation functions are used after each hidden layer and a Softplus activation function is applied after the output layer to ensure the 1D saturated potential outputs are positive and continuous. A band pass filter is used at very low ($k_y/k_{max} = 0.04$) and high ($k_y/k_{max} = 2.5$) indices of $k_y$ to prevent spikes when tested with GFS linear inputs (discussed in section \ref{sec_gfs_flux}), which exhibit mode transitions. The output of the neural network is the normalized 1D saturated potential $\phi^2_{k_y}/\Delta k_y$.

\subsection{Loss function} 
The expression for the loss function $L$ is:
\begin{align}
    L = \left |\left|\dfrac{\phi^{{NN}^2}_{k_y}}{\Delta k_y}
     - \dfrac{\phi^{{NL}^2}_{k_y}}{\Delta k_y}\right |\right|^2
     + \left |\left|2W^L_i\dfrac{\phi^{{NN}^2}_{k_y}}{\Delta k_y}\Delta k_y - \dfrac{Q_i}{\Lambda_i}\right |\right|^2
     + \left |\left|2W^L_e\dfrac{\phi^{{NN}^2}_{k_y}}{\Delta k_y}\Delta k_y - \dfrac{Q_e}{\Lambda_e}\right|\right|^2
     +\left|\left|2W^L_p\dfrac{\phi^{{NN}^2}_{k_y}}{\Delta k_y}\Delta k_y- \dfrac{\Gamma}{\Lambda_p}\right|\right|^2
\end{align} \label{loss}
The first term of the loss function is the mean squared loss between the normalized 1D saturated potential obtained from the neural network and the nonlinear data. The subsequent terms in the loss function are also mean squared error terms which incorporate species information as well from the total flux equation given in equation \ref{ql3}. Modifying the loss function this way helps in creating a physics guided neural network which promotes the satisfaction of the total flux equation for each of the species. Dividing the flux $Q$ by the quasilinear approximation function $\Lambda$, helps in separating the linear and nonlinear quantities ensuring the combination of the linear inputs and neural network outputs are able to minimise the losses with respect to the nonlinear quantities.

\subsection{Normalization of inputs and outputs} \label{ML1}
Normalizing the inputs and outputs helps in scaling the dataset, leading to an optimal mapping from the input to the output space as well as faster convergence times. The inputs are normalized as follows. The binormal wavenumber $k_y$ is normalized by $k_{max}$ giving $k_y \rightarrow \dfrac{k_y}{k_{max}}$, the growth rate $\gamma$ by $\gamma_{max}$ giving $\gamma\rightarrow\dfrac{\gamma}{\gamma_{max}}$ and the frequency by $\gamma_{max}$ giving $\omega\rightarrow \dfrac{\omega}{\gamma_{max}}$. The dimensionally-consistent combination of linear quantities for the 1D squared saturation potential is $\propto B^2_{unit}\gamma^2_{max}/k^4_{max}$. This form is derived from balance calculations between linear growth and turbulent $E \times B$ advection \cite{lee1984nonlinear}. Hence, $\phi^2_{k_y}$ is normalized by $\dfrac{B^2_{unit}\gamma_{max}^2}{k_{max}^{4}}$ to be dimensionally consistent, giving the target $\dfrac{\phi^2_{k_y}}{\Delta k_y}  \rightarrow \dfrac{\phi^2_{k_y}/(B^2_{unit}\gamma_{max}^2/k_{max}^{4})}{\Delta {k_y}/k_{max}} \rightarrow \dfrac{\phi^2_{k_y}}{\Delta k_y}\bigg{/}\dfrac{B^2_{unit}\gamma_{max}^2}{k_{max}^{5}}$. The total flux $Q$, when used in the loss function \ref{loss}, is normalized by $\dfrac{B^2_{unit}\gamma_{max}^2}{k_{max}^{3}}$ to maintain dimensional consistency across the flux equation \ref{ql3}.

\subsection{Training the model}
Because the NL CGYRO dataset is discretized to a high binormal resolution, splitting it randomly into separate training and testing sets over the binormal wavenumber $k_y$ would still preserve the dominant features even at lower resolutions, and therefore would not indicate a good test of network generalization. A better test of the neural network model would be to train with a dataset that excludes entire parameter scans, and then later test the trained model on those scans that were left out during training to assess the generalizability of the model, which is explored in section \ref{sec_left_flux}. Therefore, as a baseline regression task, the entire dataset of 43 cases was used for training the neural network model. 

\section{Results}
\subsection{1D Saturated Potentials}\label{sec_cgyro_pot}
\begin{figure}[!ht]
\centering
    \begin{subfigure}[b]{0.32\textwidth}
    \centering
    \includegraphics[width=\textwidth]{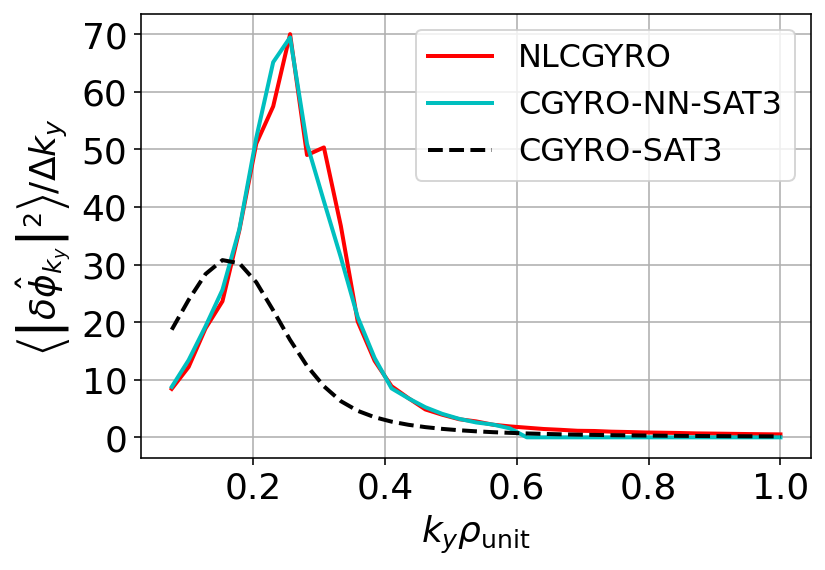}
    \caption{{\scriptsize $a/L_T = 1.5$, D}}
    \label{prof1}
    \end{subfigure}
    \begin{subfigure}[b]{0.32\textwidth}
    \centering
    \includegraphics[width=\textwidth]{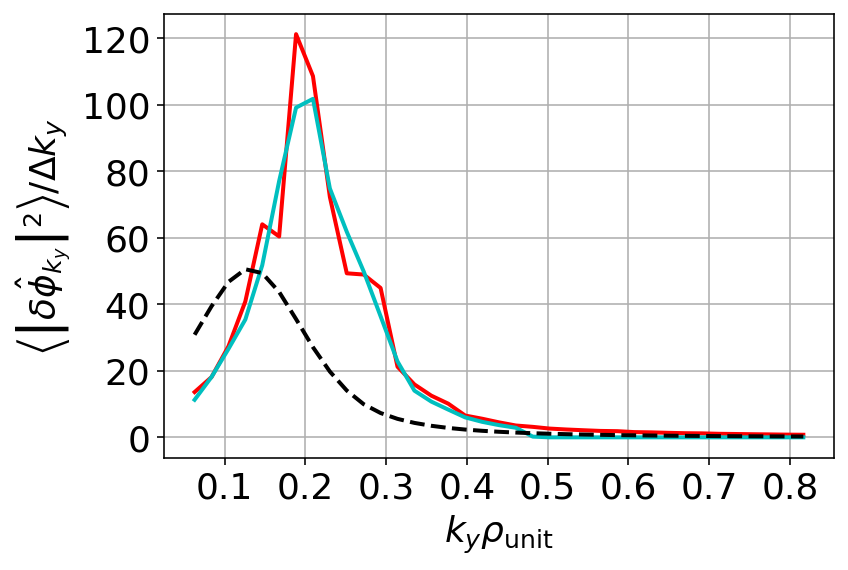}
    \caption{{\scriptsize$a/L_T = 1.5$, T}}
    \label{prof2}
    \end{subfigure}
    \begin{subfigure}[b]{0.32\textwidth}
    \centering
    \includegraphics[width=\textwidth]{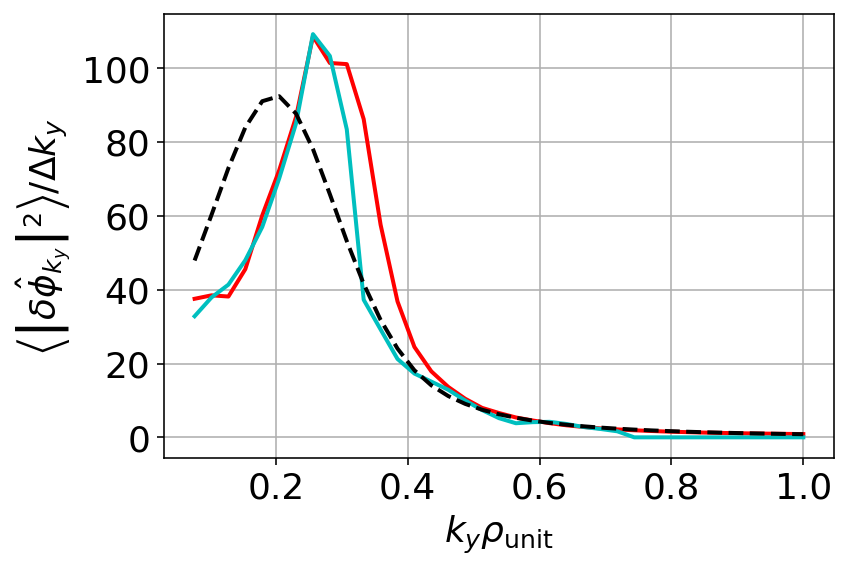} 
    \caption{{\scriptsize$(a/c_s)\nu_{ee} = 1.0$, $a/L_n = 3.0$, D}}
    \label{prof3}
    \end{subfigure}
    \begin{subfigure}[b]{0.32\textwidth}
    \centering
    \includegraphics[width=\textwidth]{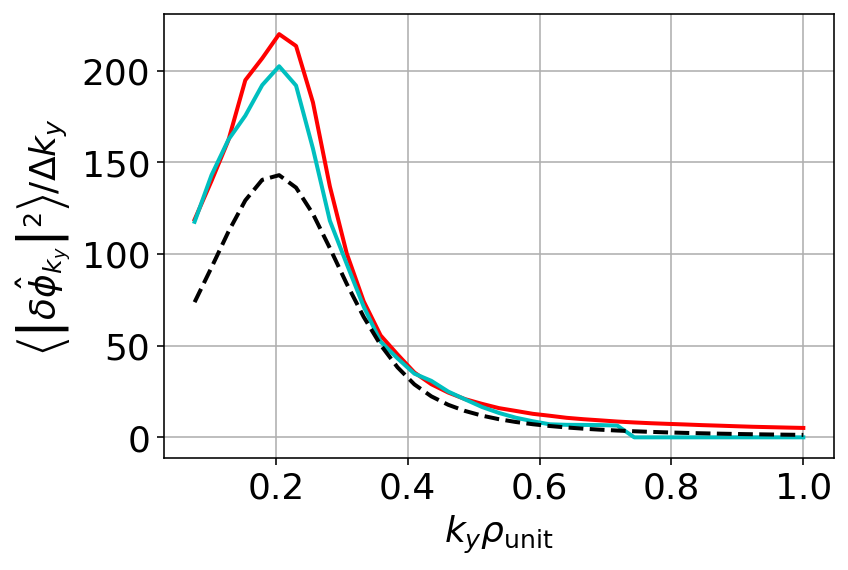}
    \caption{{\scriptsize$\hat{s} = 0.5$, D}}
    \label{prof4}
    \end{subfigure}
    \begin{subfigure}[b]{0.32\textwidth}
    \centering
    \includegraphics[width=\textwidth]{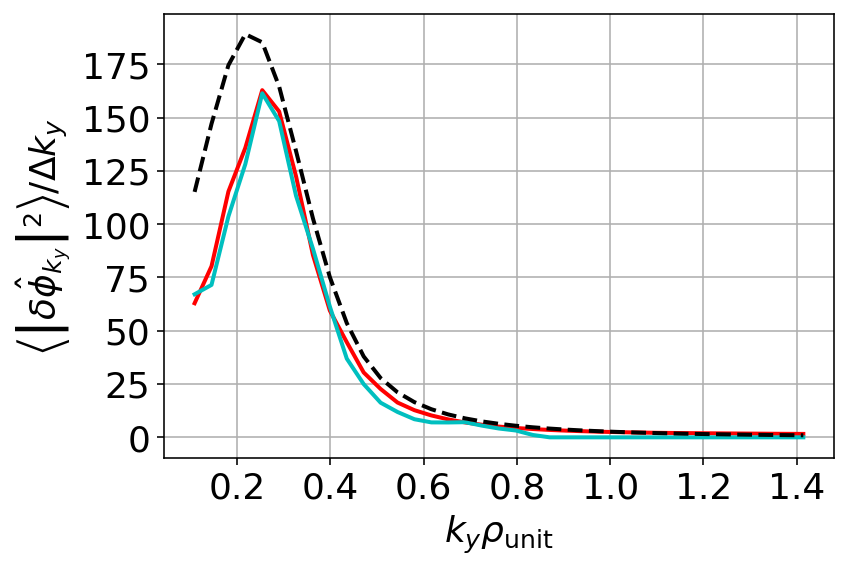}
    \caption{{\scriptsize$a/L_n = 2.0$, H}}
    \label{prof5}
    \end{subfigure}
    \begin{subfigure}[b]{0.32\textwidth}
    \centering
    \includegraphics[width=\textwidth]{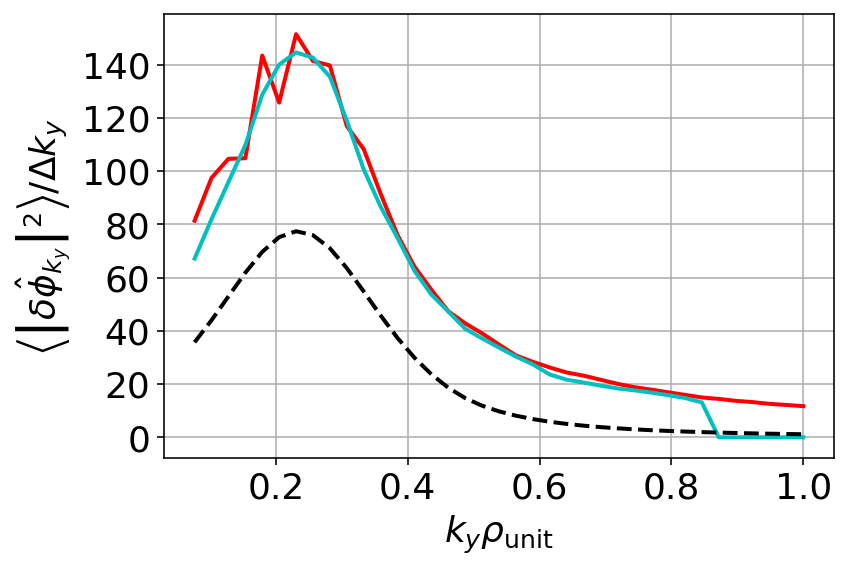}
    \caption{{\scriptsize$\Delta = -0.5$, D}}
    \label{prof6}
    \end{subfigure}
    \begin{subfigure}[b]{0.32\textwidth}
    \centering
    \includegraphics[width=\textwidth]{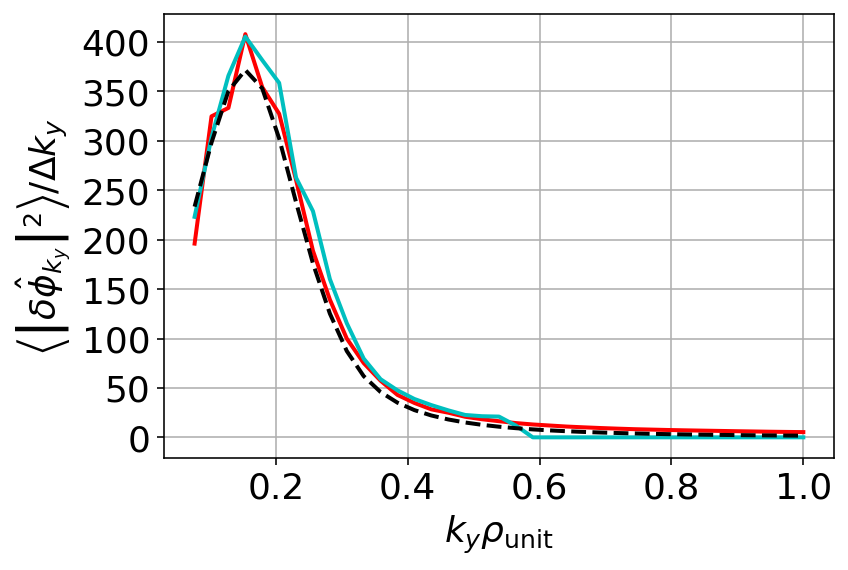}
    \caption{{\scriptsize GA-std}}
    \label{prof7}
    \end{subfigure}
    \begin{subfigure}[b]{0.32\textwidth}
    \centering
    \includegraphics[width=\textwidth]{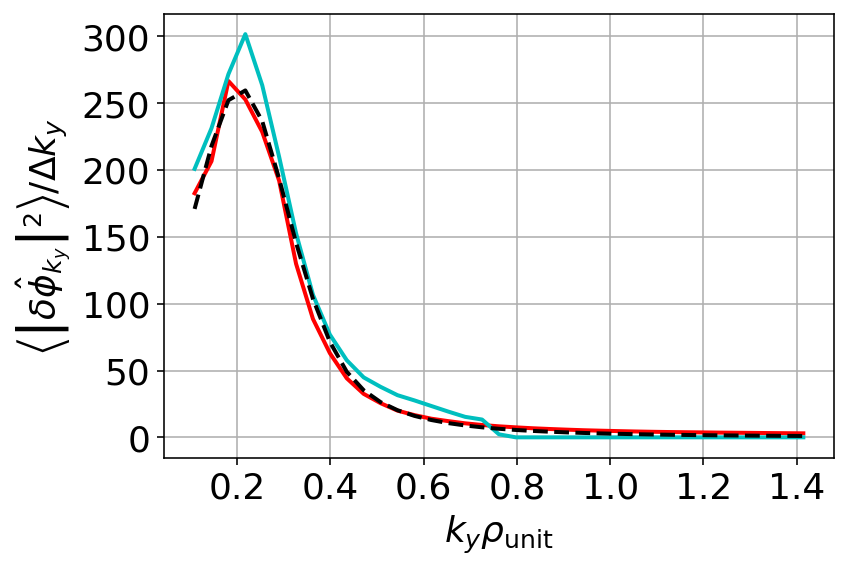}
    \caption{{\scriptsize$a/L_T = 3.5$, H}}
    \label{prof8}
    \end{subfigure}
    \begin{subfigure}[b]{0.32\textwidth}
    \centering
    \includegraphics[width=\textwidth]{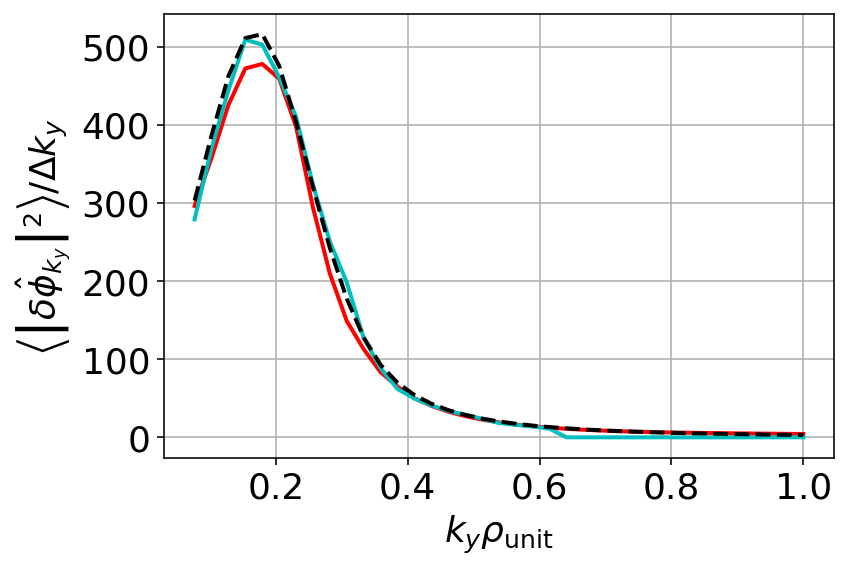}
    \caption{{\scriptsize$a/L_n = 3.0$, D}}
    \label{prof9}
    \end{subfigure}
    \captionsetup{justification=raggedright,singlelinecheck=false}
    \caption{1D saturated potentials for NL CGYRO (solid red), CGYRO-SAT3 (black dashed) and CGYRO-NN-SAT3 (solid cyan) against binormal wavenumber $k_y$.}
    \label{prof}
\end{figure}

Figure \ref{prof} shows a selection of the spectra of the 1D saturated potentials plotted against binormal wavenumber $k_y$. Each case corresponds to one of the 43 scans from the simulation database, with a subset shown here for comparison. The labels of each plot mention the parameter scan, its value and H, D, T refer to the isotopes. A general qualitative improvement in the mappings of the saturated potentials from linear CGYRO data is seen in the NN-SAT3 model compared to the SAT3 model. In Figure \ref{prof}, the first row with figures \ref{prof1}, \ref{prof2}, \ref{prof3}, show cases where the peak locations as well as the peak heights of the 1D saturated potentials are mapped more accurately by CGYRO-NN-SAT3. The middle row, consisting of figures \ref{prof4}, \ref{prof5}, \ref{prof6}, show those cases where the peak heights are mapped more accurately by the neural network model. Cases where CGYRO-SAT3 both over- and under-predicts the NL CGYRO 1D saturated potentials are shown in contrast to CGYRO-NN-SAT3 which is able to map more accurately. And the bottom row, with figures \ref{prof7}, \ref{prof8}, \ref{prof9}, show cases where CGYRO-NN-SAT3 performs as good as the CGYRO-SAT3. 

\begin{figure}[!ht]
\centering
    \begin{subfigure}[b]{0.49\textwidth}
    \centering
    \includegraphics[width=\textwidth]{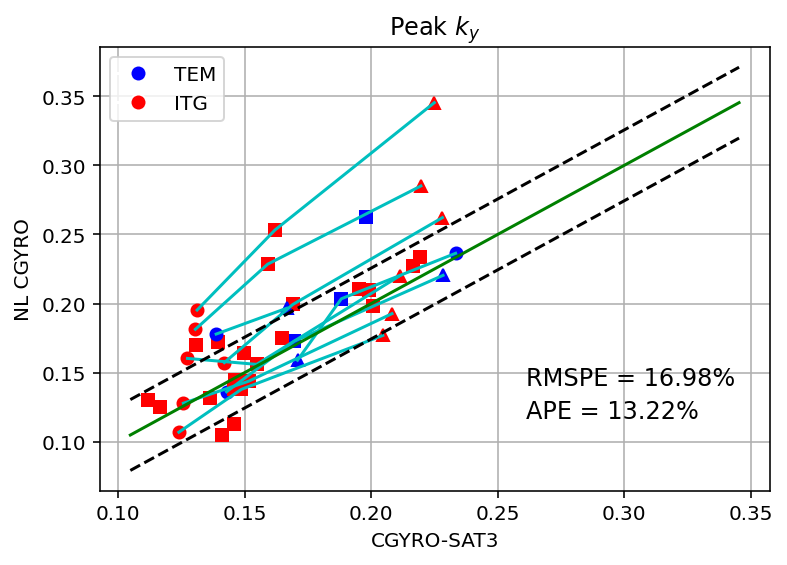}
    \caption{Peak locations CGYRO-SAT3 vs NL CGYRO}
    \label{sum1}
    \end{subfigure}
    \hfill
    \begin{subfigure}[b]{0.49\textwidth}
    \centering
    \includegraphics[width=\textwidth]{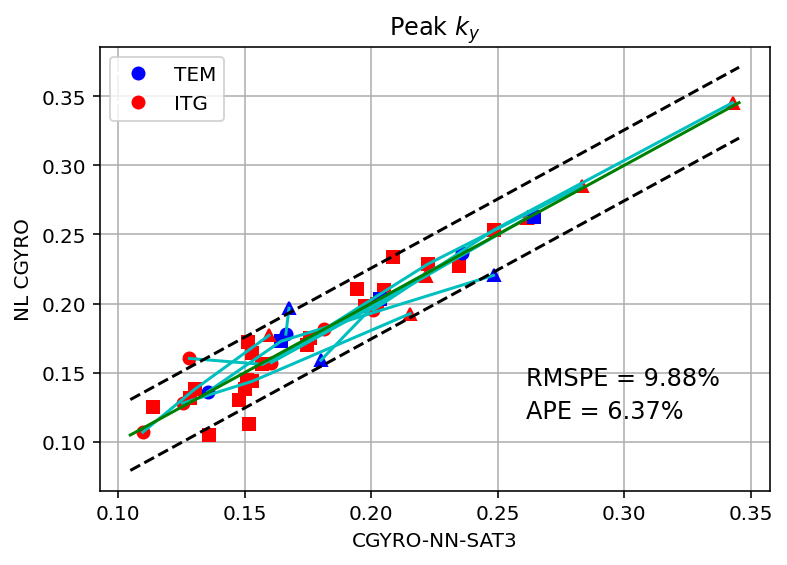}
    \caption{Peak locations CGYRO-NN-SAT3 vs NL CGYRO}
    \label{sum2}
    \end{subfigure}
    \begin{subfigure}[b]{0.49\textwidth}
    \centering
    \includegraphics[width=\textwidth]{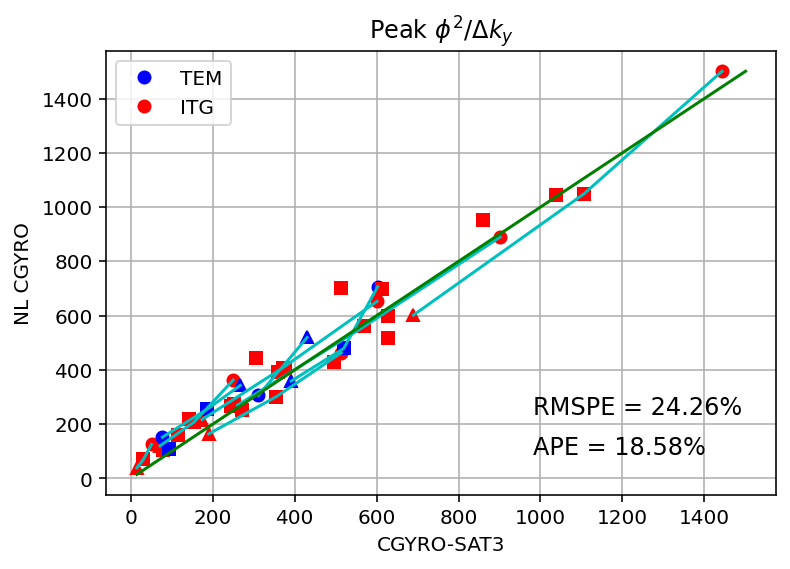} 
    \caption{Peak $\phi^2/\Delta k_y$ CGYRO-SAT3 vs NL CGYRO}
    \label{sum3}
    \end{subfigure}
    \hfill
    \begin{subfigure}[b]{0.503\textwidth}
    \centering
    \includegraphics[width=\textwidth]{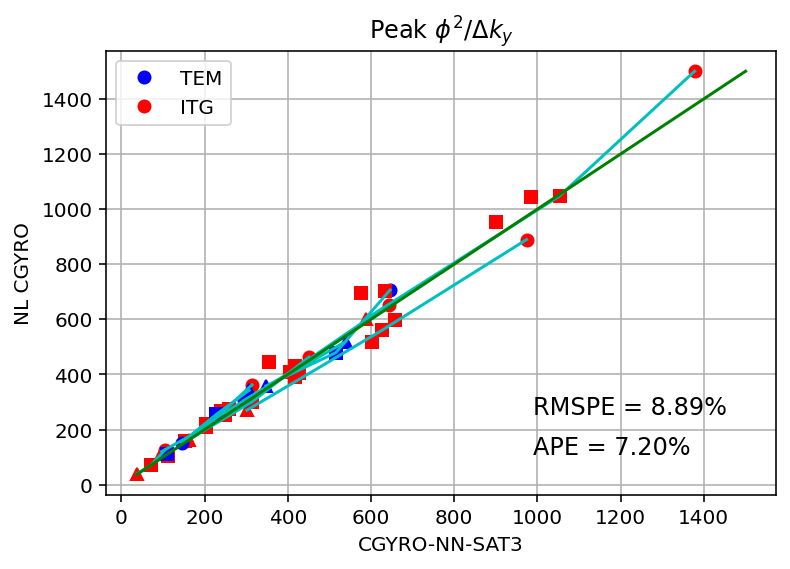}
    \caption{Peak $\phi^2/\Delta k_y$ CGYRO-NN-SAT3 vs NL CGYRO}
    \label{sum4}
    \end{subfigure}
    \captionsetup{justification=raggedright,singlelinecheck=false}
    \caption{Summary plots comparing the peak values, shown in subplots (a) and (b), and peak locations, shown in (c) and (d), of the 1D saturated potentials for CGYRO-SAT3 and CGYRO-NN-SAT3. The two turbulence modes are shown in red (ITG) and blue (TEM). The isotopes markers are triangle (Hydrogen), square (Deuterium) and circle (Tritium). Black dashed lines in (a) and (b) show a band of $\pm \Delta k_{y_D}$. Errors shown in APE (Average Percentage Error) and RMSPE (Root Mean Squared Percentage Error).}
    \label{summary}
\end{figure}

To quantify the improvement in the mappings of the 1D saturated potentials using the CGYRO-NN-SAT3 over CGYRO-SAT3, summary plots are created of all the $43$ cases for the peak location and peak values. Figures \ref{sum1} and \ref{sum2} show the summary plots for the location of the spectrum peak in $k_y$ of the 1D saturated potentials for the SAT3 and the NN-SAT3 models respectively. Black dashed lines show a band width of $\pm \Delta k_{y_D}$, the normalized grid spacing, which takes into account the grid discretization while calculating the peak location of the saturated potential in the grid. The peak location mappings in CGYRO-NN-SAT3 show some improvement over CGYRO-SAT3 with the root mean squared percentage error (RMSPE) decreasing from 16.98\% to 9.88\% and average percentage error (APE) decreasing from 13.22\% to 6.37\%. The low ($a/L_T = 1.5$) and medium ($a/L_T = 2.25$) temperature gradient ITG cases, shown in red markers, lie outside the $\pm \Delta k_{y_D}$ band in CGYRO-SAT3, whereas for CGYRO-NN-SAT3 they lie inside the band. The peak values of the 1D saturated potentials are shown in figures \ref{sum3} and \ref{sum4} for CGYRO-SAT3 and CGYRO-NN-SAT3 respectively. The mapping of the peak values show a marked improvement in CGYRO-NN-SAT3 with an RMSPE of 8.89\% and APE of 7.20\% as compared to CGYRO-SAT3 with an RMSPE of 24.26\% and APE of 18.58\%. 
\subsection{Fluxes}\label{sec_cgyro_flux}
\begin{figure}[!ht]
    \centering
        \includegraphics[width=1.0\linewidth]{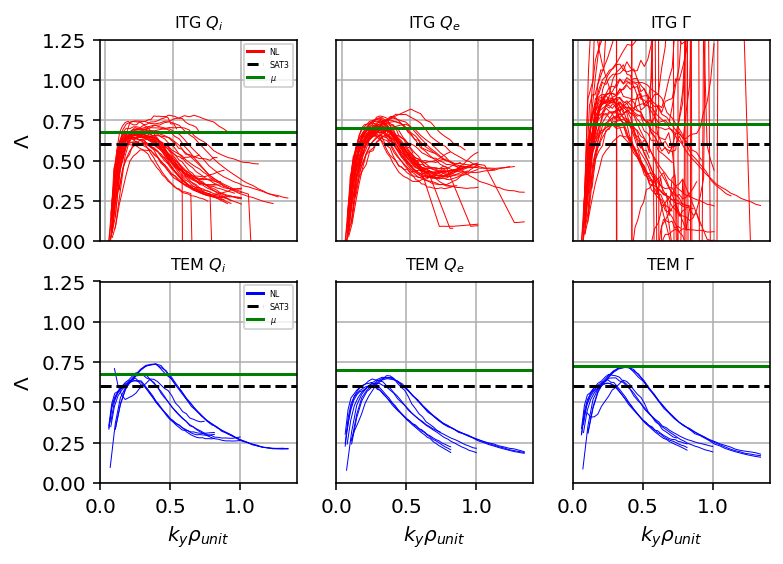}
    \captionsetup{justification=raggedright,singlelinecheck=false}
    \caption{Quasilinear Approximation Function for CGYRO (solid red for ITG and solid blue for TEM), SAT3 (black dashed) and  the mean of the maximum QLA value of each case (green)}
    \label{qla}
\end{figure}

 Having obtained an improved mapping of the 1D saturated potential, we now consider the prediction of the total fluxes. The values of the 1D saturated potential obtained from the neural network output are unnormalized and then plugged into equation \ref{ql3} in order to obtain the final species flux in each channel, i.e., ion energy flux $Q_i$, electron energy flux $Q_e$ and particle flux $\Gamma$. A plot showing the quasilinear approximation (QLA) function $\Lambda$ obtained from the CGYRO dataset is shown in figure \ref{qla}. The QLA values obtained from the CGYRO runs are shown in red for the ITG cases and in blue for the TEM cases. From left to right are the QLA values for  $Q_i$,  $Q_e$ and  $\Gamma$ respectively. As can be observed, the particle fluxes for the ITG case is widely varying as compared to the other fluxes. For the SAT3 model, shown in black dashed lines, a mode dependent transition function, one each for the heat energy and particle fluxes is used where $\Lambda^{\Gamma}_{ITG} = 1.1$, $\Lambda^{\Gamma}_{TEM} = \Lambda^{Q}_{TEM} = 0.6$ and $\Lambda^{Q}_{ITG} = 0.75$. For this work, the maximum value of the CGYRO QLA values is calculated for all the 43 parameter scans and then the mean is calculated. This mean value of $\Lambda = 0.72$, shown in green, is used as the constant QLA value in the total flux equation \ref{ql3} and was found to fit well for both ITG and TEM cases as well as for all the fluxes in each channel. Hence, a single constant value of the QLA could be applied to all cases for the NN-SAT3 model, which unlike the SAT3 model, precludes the need for a separate classification of the QLA for mode and species. 

\begin{figure}[!ht]
\centering
    \begin{subfigure}[b]{0.32\textwidth}
    \centering
        \includegraphics[width=\textwidth]{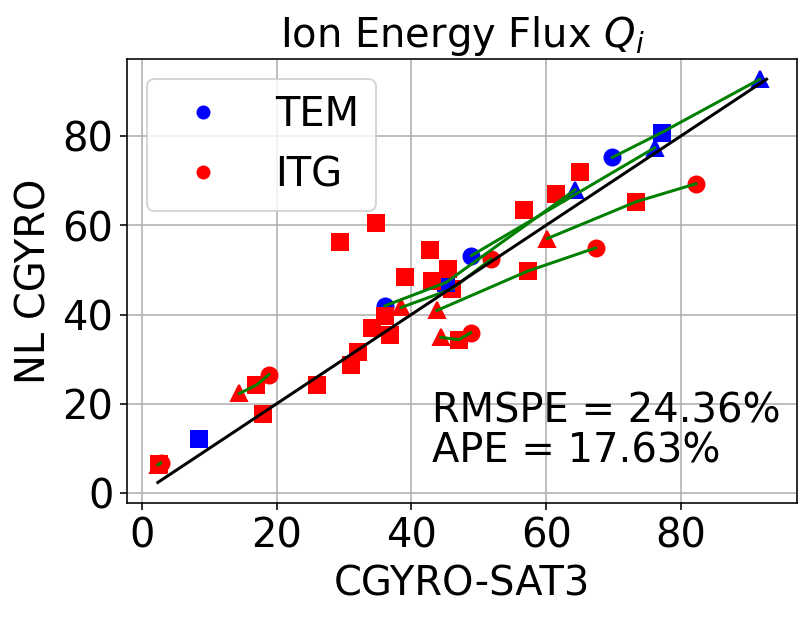}
        \caption{}
        \label{flux_qi_s3}
    \end{subfigure}
    \begin{subfigure}[b]{0.32\textwidth}
    \centering
        \includegraphics[width=\textwidth]{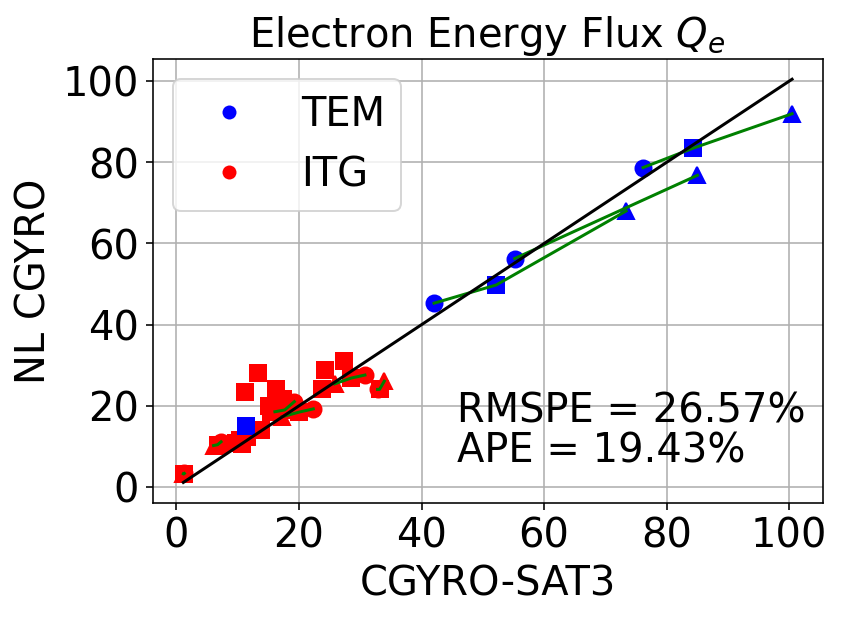}
        \caption{}
        \label{flux_qe_s3}
    \end{subfigure}
    \begin{subfigure}[b]{0.32\textwidth}
    \centering
        \includegraphics[width=\textwidth]{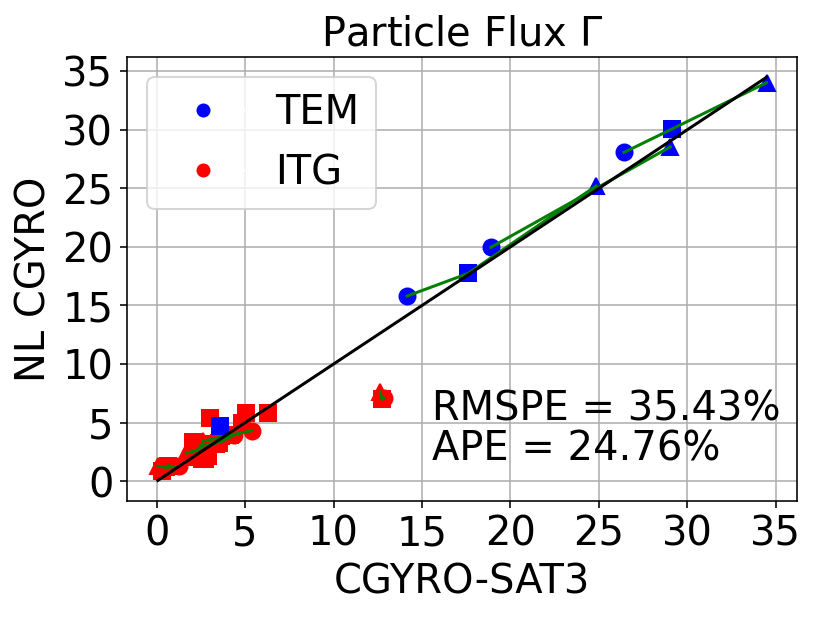}
        \caption{}
        \label{flux_qp_s3}
    \end{subfigure}
    \begin{subfigure}[b]{0.32\textwidth}
    \centering
        \includegraphics[width=\textwidth]{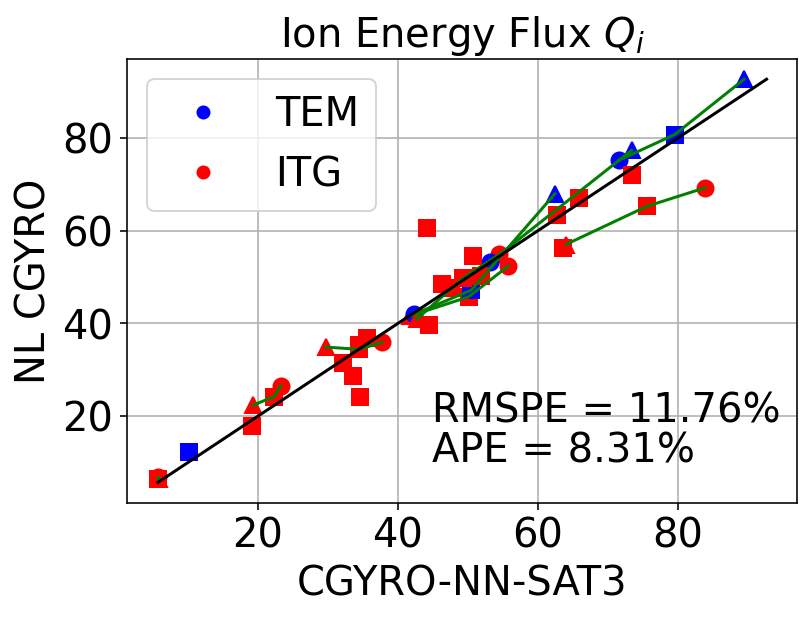}
        \caption{}
        \label{flux_qi_nns3}
    \end{subfigure}
    \begin{subfigure}[b]{0.32\textwidth}
    \centering
        \includegraphics[width=\textwidth]{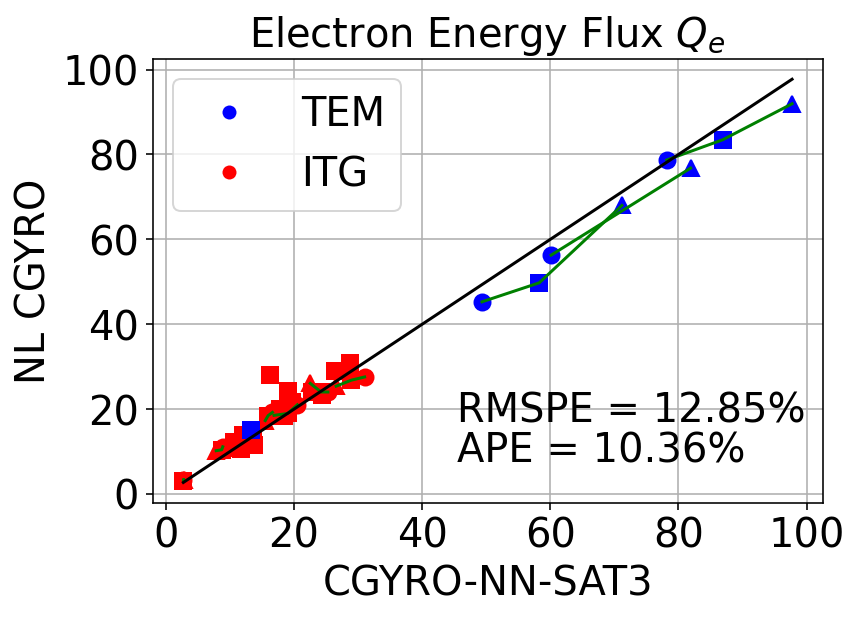}
        \caption{}
        \label{flux_qe_nns3}
    \end{subfigure}
    \begin{subfigure}[b]{0.32\textwidth}
    \centering
        \includegraphics[width=\textwidth]{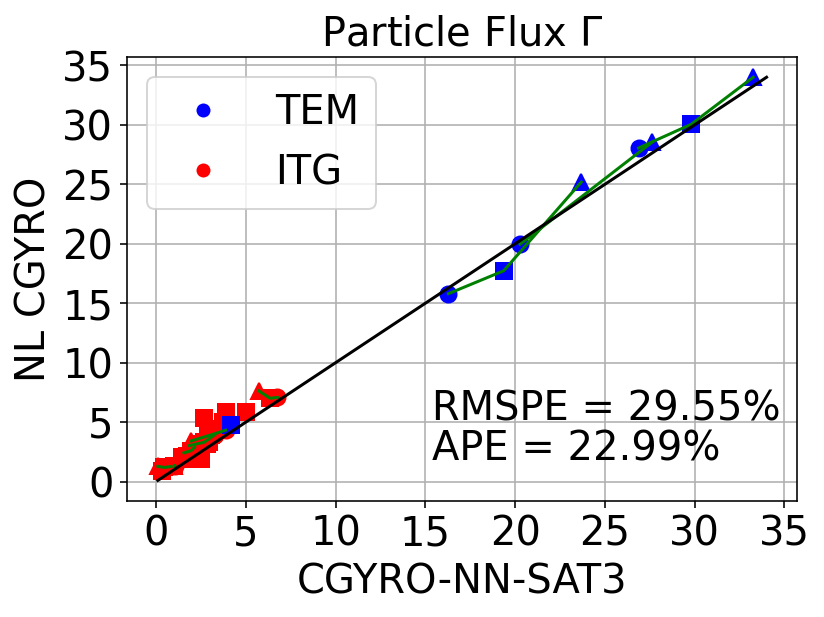}
        \caption{}
        \label{flux_qp_nns3}
    \end{subfigure}
    \captionsetup{justification=raggedright,singlelinecheck=false}
    \caption{Comparison of fluxes obtained from CGYRO-SAT3 and CGYRO-NN-SAT3 plotted against NL CGYRO data. The two turbulence modes are shown in red (ITG) and blue (TEM). The isotopes markers are triangle (Hydrogen), square (Deuterium) and circle (Tritium). Errors shown in APE and RMSPE.}
    \label{flux}
\end{figure}

The flux obtained using these quantities is then compared with the NL CGYRO flux. Figure \ref{flux} shows a comparison of $Q_i$ , $Q_e$ and $\Gamma$ obtained from CGYRO-SAT3 and CGYRO-NN-SAT3, plotted against the NL CGYRO flux. For the cases with two or three isotopes, CGYRO-SAT3 is able to capture the isotope scaling, which is the scaling of flux magnitudes of the isotopes with a unit slope. In CGYRO-SAT1, which is developed using a database containing only Deuterium cases, the lines connecting the markers representing the isotopes would be almost perpendicular to the identity line \cite{dudding2022new}. In contrast, CGYRO-SAT3, shown in figures \ref{flux_qi_s3}, \ref{flux_qe_s3}, \ref{flux_qp_s3}, is reasonably able to capture isotope scaling with the straight line connecting the isotope markers being almost parallel and much closer to the identity line. Similar plots are made for the fluxes obtained using CGYRO-NN-SAT3, as shown in figures \ref{flux_qi_nns3}, \ref{flux_qe_nns3}, \ref{flux_qp_nns3}, where it can be seen that CGYRO-NN-SAT3 is also able to model isotope scaling.

For CGYRO-NN-SAT3, a large reduction is observed in the APE and RMSPE over CGYRO-SAT3 for $Q_i$ and $Q_e$. For $Q_i$, the APE is down from 17.63\% to 8.31\% and the RMSPE is down from 24.36\% to 11.76\%. For $Q_e$, the APE is down from 19.43\% to 10.36\% and the RMSPE is down from 26.57\% to 12.85\%. For $\Gamma$, there is some improvement in the APE from 24.76\% to 22.99\% and RMSPE from 35.43\% to 29.55\% of the CGYRO-NN-SAT3 over CGYRO-SAT3. In particular, the scan with density gradient length scale $a/L_n = 2.0$ at the highest ITG particle fluxes is mapped better with CGYRO-NN-SAT3 than CGYRO-SAT3. The relatively smaller improvement in the overall particle flux for the neural network model can be attributed to the fact that we use one single constant QLA value for both the ITG and TEM cases. Whereas the QLA profile values for the particle flux obtained from NL CGYRO, as seen in figure \ref{qla}, show a wide range of variation for ITG cases. 

\begin{figure}[!ht]
\centering
    \begin{subfigure}[b]{0.49\textwidth}
        \centering
        \includegraphics[width=\textwidth]{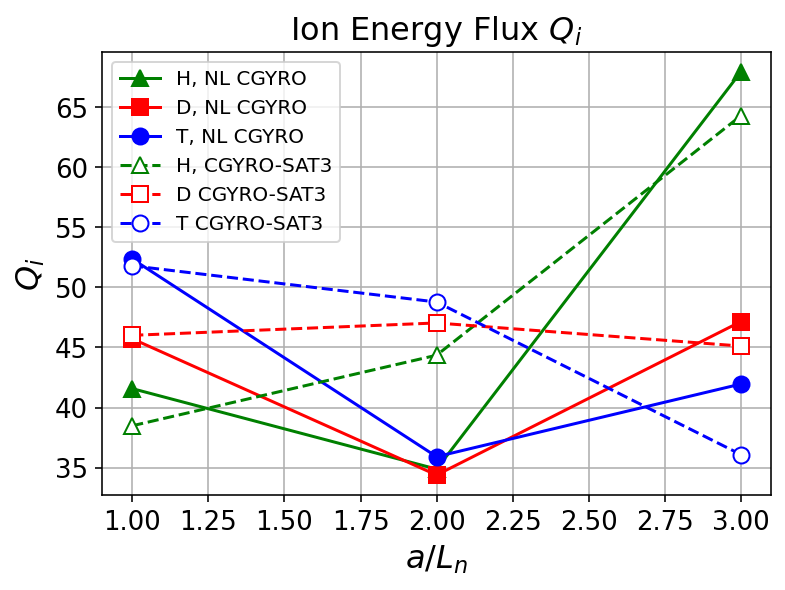}
        \caption{}
        \label{flux_aln_s3}
    \end{subfigure}
    \begin{subfigure}[b]{0.49\textwidth}
        \centering
        \includegraphics[width=\textwidth]{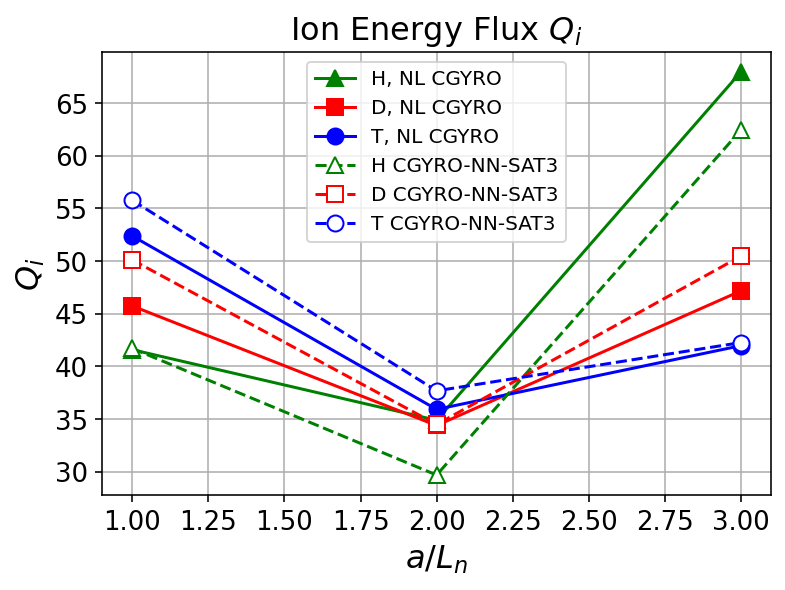}
        \caption{}
        \label{flux_aln_nns3}
    \end{subfigure}
    \begin{subfigure}[b]{0.49\textwidth}
        \centering
        \includegraphics[width=\textwidth]{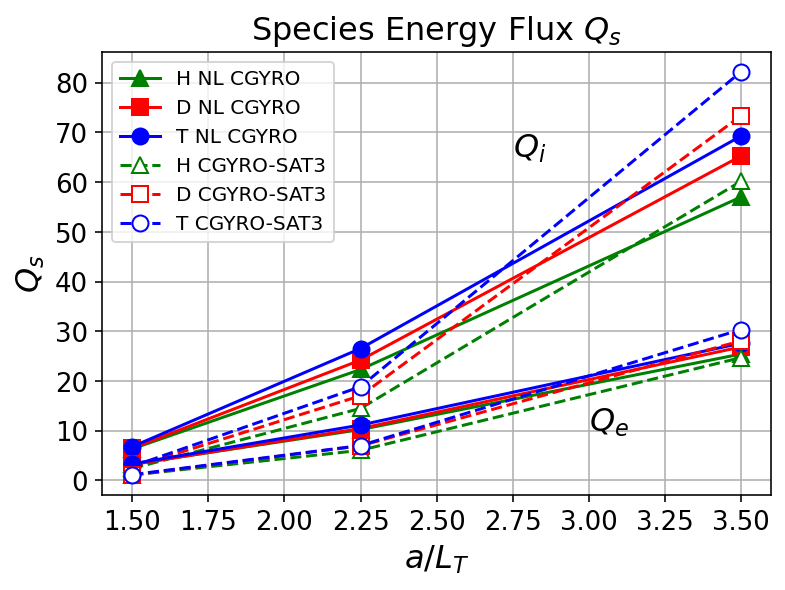}
        \caption{}
        \label{flux_alt_s3}
    \end{subfigure}
    \begin{subfigure}[b]{0.49\textwidth}
        \centering
        \includegraphics[width=\textwidth]{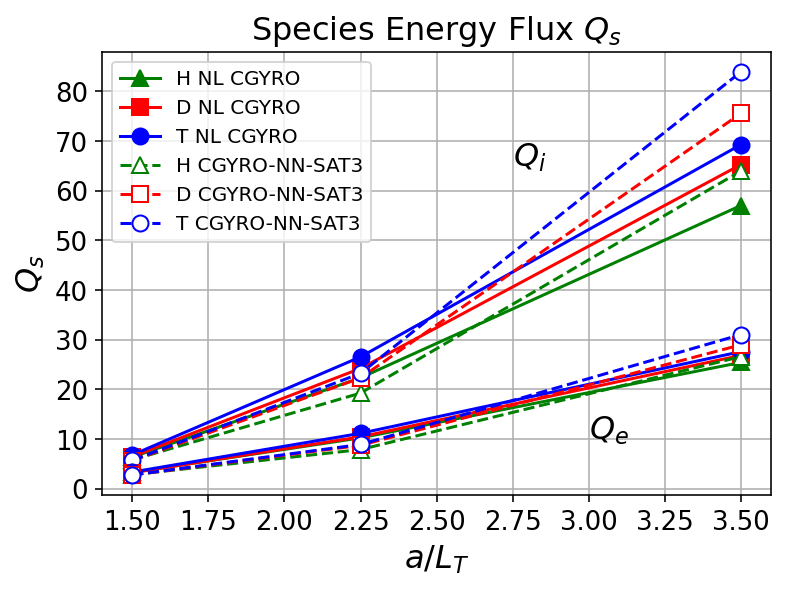}
        \caption{}
        \label{flux_alt_nns3}
    \end{subfigure}
    \captionsetup{justification=raggedright,singlelinecheck=false}
    \caption{Ion energy fluxes for NL CGYRO (solid lines) and CGYRO-SAT3 (dashed lines) (a), CGYRO-NN-SAT3 (dashed lines) (b) against density gradient scale length $a/L_n$. Ion and electron energy fluxes for NL CGYRO (solid lines) and CGYRO-SAT3 (dashed lines) against temperature gradient scale length, $a/L_{T_i} = a/L_{T_e}$ (c), CGYRO-NN-SAT3 (dashed lines) (d) against temperature gradient scale length, $a/L_{T_i} = a/L_{T_e}$. The isotopes markers are triangle (Hydrogen), square (Deuterium) and circle (Tritium).}
    \label{fluxq}
\end{figure}

A selection of energy flux scans with various key tokamak parameters are shown in figure \ref{fluxq}, where CGYRO-SAT3 and CGYRO-NN-SAT3 are compared by plotting them against NL CGYRO data. The density gradient scan is shown in figures \ref{flux_aln_s3} and \ref{flux_aln_nns3}. Figure \ref{flux_aln_s3} shows the comparison of CGYRO-SAT3 (dashed lines) against NL CGYRO (solid lines), and figure \ref{flux_aln_nns3} shows CGYRO-NN-SAT3 (dashed lines) against NL CGYRO (solid lines). These plots demonstrate the positive isotope scaling for the ITG-dominated GA-std cases at low density gradient $a/L_n = 1.0$, the grouping of the fluxes for the transition $a/L_n = 2.0$ case, and the anti-gyroBohm scaling at the high density gradient TEM-dominated $a/L_n = 3.0$ case. Moreover, CGYRO-NN-SAT3 is able to capture the fluxes more accurately for density gradient $a/L_n = 2.0$ than CGYRO-SAT3, which is because of the neural network model's ability to capture the peak heights of the 1D saturated potential far more accurately for this parameter scan, as shown in figure \ref{prof5} in section \ref{sec_cgyro_pot} and later in figures \ref{gfs5} and \ref{gfs6} in section \ref{sec_gfs_flux}.
 
The ion and electron energy heat fluxes against matched temperature gradients $(a/L_{T_i} = a/L_{T_e})$ are shown in figure \ref{flux_alt_s3} for CGYRO-SAT3 and figure \ref{flux_alt_nns3} for CGYRO-NN-SAT3. The general trend in both isotope scaling and magnitude can be observed to agree with the NL CGYRO data. The low $a/L_T = 1.5$ and medium $a/L_T = 2.25$ temperature gradient scans, are mapped better by CGYRO-NN-SAT3 than by CGYRO-SAT3, capturing the peak locations and peak values of the 1D saturated potentials more accurately. For CGYRO-SAT3 in figure \ref{flux_alt_s3}, the fluxes appear to be somewhat under-predicted near the point of the critical temperature gradient. In this region, which is a key region of parameter space for experimental conditions, the heat fluxes being very small, can be easily over- or under-predicted by saturation rules \cite{dudding2026integrated}. Since the turbulent transport is stiff at the regions of low temperature gradient, CGYRO-NN-SAT3 may yield improved predictions in integrated modeling due to more accurately finding the fluxes just above the critical gradient.

\subsection{Model accuracy with parameters left out during training}\label{sec_left_flux}
\begin{figure}[!ht]
\centering
    \includegraphics[width=1.0\textwidth]{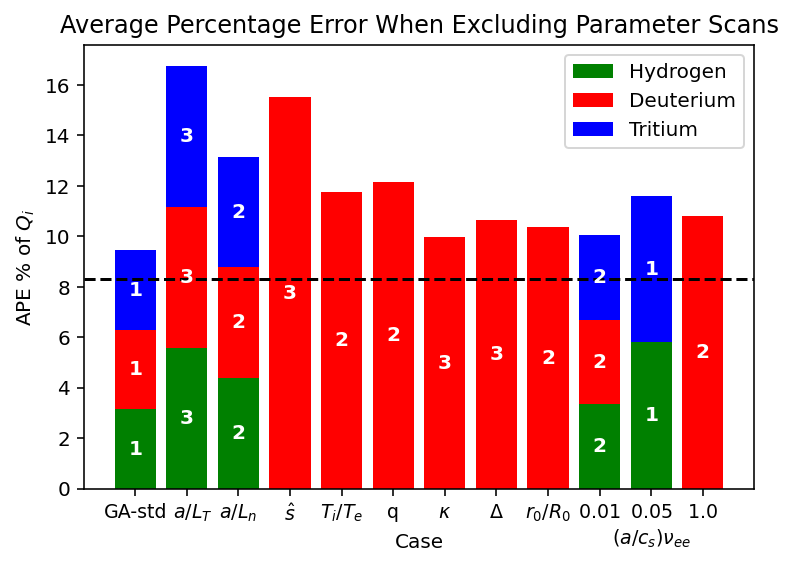} 
    \captionsetup{justification=raggedright,singlelinecheck=false}
    \caption{Average percentage errors in ion energy flux when removing individual parameter scans. Green bars correspond to Hydrogen, red to Deuterium, blue to Tritium. The number inside each colored bar corresponds to the number of cases in that isotope. Black dashed line is the baseline average percentage error.}
    \label{bar}
\end{figure}

In the preceding results for the 1D saturated potentials and fluxes obtained from CGYRO-NN-SAT3, the entire dataset was used for training. To test the ability of the model to predict when tested on new data, the performance of the network is evaluated on test cases that are not included in the training dataset, by training the neural network with a certain parameter scan left out during training, and then testing the neural network on the entire dataset. The bar plot shown in figure \ref{bar} shows APE in $Q_i$ of the entire network when a particular parameter scan, as represented by a bar, is left out during training. The height of the bar represents the total APE across all the 43 simulations. Each isotope is color coded and stacked in the bars, with Hydrogen cases shown in green, Deuterium shown in red and Tritium shown in blue. The number inside each colored stacked bar designates the number of test cases in that isotope corresponding to the parameter scan on the x-axis. As can be seen from the plot, certain parameter scans have more test cases than others. For example the scan for temperature gradient consists of 9 cases (3 each for Hydrogen, Deuterium and Tritium), whereas the scan for the ratio of minor and major radius consists of only 2 cases (both Deuterium). The black dashed line is the baseline APE in $Q_i$ when the entire dataset is used for training. The heights of all the bars are above the baseline APE, indicating a deterioration of model performance when the cases corresponding to those bars are left out during the training process. The deterioration is worse for some cases than others. For example, the errors are highest when the scans for density gradient $a/L_n$, temperature gradient $a/L_T$ and magnetic shear $\hat{s}$ are left out during training, thus indicating a stronger contribution to overall network performance by those parameter scans. In contrast, the bars corresponding to other cases which are slightly above the baseline, such as the GA-std case, elongation $\kappa$ and the case with collisionality $ (a/c_s)\nu_{ee} = 0.01$, do not play a major role in the overall accuracy of the network. This knowledge will help in guiding us towards decisions about future high fidelity parameters scans when we expand our dataset.

\begin{figure}[!ht]
\centering
    \begin{subfigure}[b]{0.32\textwidth}
        \centering
        \includegraphics[width=1.0\textwidth]{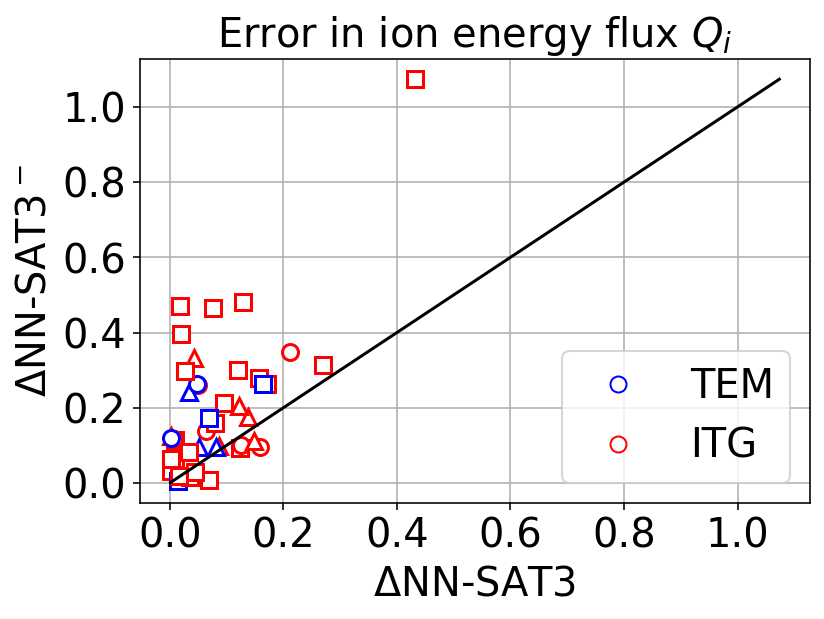}
        \caption{}
        \label{del1}
    \end{subfigure}
     \begin{subfigure}[b]{0.32\textwidth}
        \centering
        \includegraphics[width=1.0\textwidth]{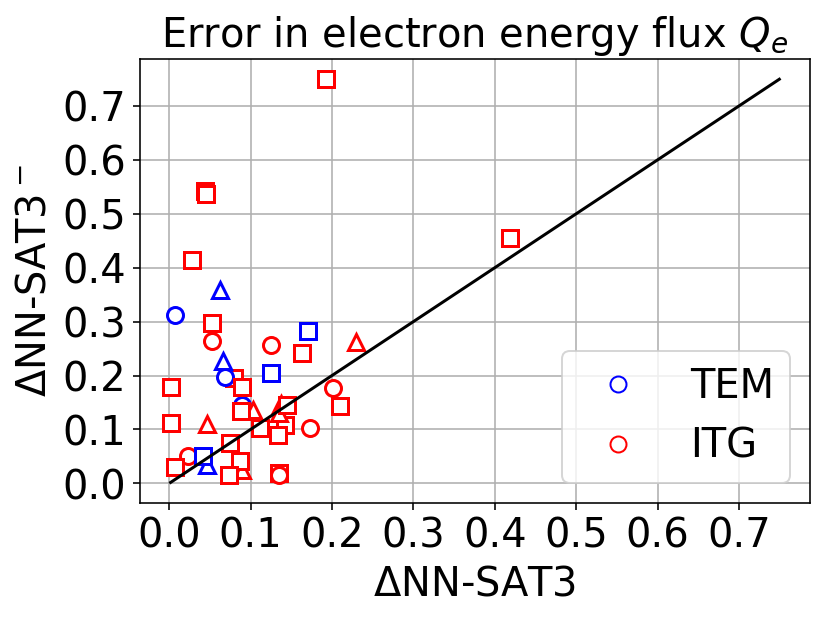}
        \caption{}
        \label{del2}
    \end{subfigure}
     \begin{subfigure}[b]{0.32\textwidth}
        \centering
        \includegraphics[width=1.0\textwidth]{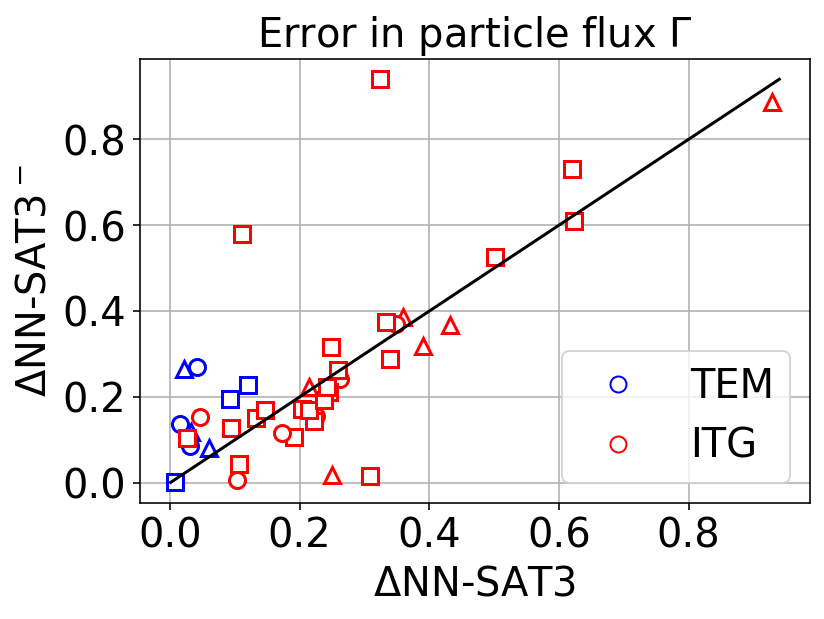}
        \caption{}
        \label{del3}
    \end{subfigure}
    \captionsetup{justification=raggedright,singlelinecheck=false}
    \caption{Error in fluxes when training with full dataset vs training with individual cases removed. The two dominant turbulence mode types are shown in red (ITG) and blue (TEM). The isotopes markers are triangle (Hydrogen), square (Deuterium) and circle (Tritium).}
    \label{del}
\end{figure}

Figure \ref{del} shows the absolute relative error in the prediction of the fluxes obtained from CGYRO-NN-SAT3 compared to the NL CGYRO fluxes for a particular case (out of 43 cases) at a fixed parameter, when that individual case is left out of the training set vs when the entire dataset is used for training. The x-axis is the absolute relative error in the predicted flux of a particular case when the entire dataset is used for training $Q_{NN}$, compared to the NL CGYRO flux $Q_{NL}$, 
\begin{align}
  \Delta \mathrm{NN\mathord{-}SAT3} = \left|\frac{Q_{NN}-Q_{NL}}{Q_{NL}}\right|
\end{align}
The y-axis is the absolute relative error in the predicted flux in the same case when that case is left out during training $Q_{NN}^-$, compared to the NL CGYRO flux using the same neural network architecture as the baseline model.
\begin{align}
    \Delta \mathrm{NN\!-\!SAT3^-}  = \left|\frac{Q_{NN}^--Q_{NL}}{Q_{NL}}\right|
\end{align} \label{L1}

 Figure \ref{del1} is the plot for the absolute relative error in $Q_i$, \ref{del2} for $Q_e$ and \ref{del3} for $\Gamma$. Each unfilled marker corresponds to a particular case which was left out during training. The color and shape of the markers follow the same convention as previous plots. In general the TEM cases, shown by blue markers, being closer to the identity line, show that their prediction accuracy is not affected much when left from the training database. In comparison, some ITG cases are far above the identity line, showing that the accuracy of the prediction of those cases is worsened when that same case is left out during training. This demonstrates the network's dependence on the presence of certain ITG cases during training to obtain an accurate fit to the NL CGYRO data. Although the neural network is able to generalize for the majority of cases, there are some cases which act like edge cases, and hence cannot be predicted via an interpolation of the existing cases, thus necessitating the need to be present in the training dataset. The markers which are above the identity line show that the neural network performs worse in their prediction when these particular cases are excluded from training, compared to when the entire dataset is used for training. Across the plots for the 3 fluxes, the marker for magnetic shear $\hat{s} = 0.25$, D is consistently above the identity line by a large extent, which is similar to what we observed in the previous section in figure \ref{bar}, where the bar height for $\hat{s}$ is significantly high above the baseline APE. Similarly, the markers which are slightly above or below the identity line show that the neural network performs just as well in predicting certain cases when those cases are left out during training, thus showing the ability of the neural network to generalize for most cases. These results give an indication of the type of parameter scans that contribute the most to the overall accuracy of the neural network model, and that future iterations of the model with an expanded training dataset should contain these parameters scanned for other values, while those parameter scans which have less of an effect on the performance of the model, can be left out or have fewer number of cases in the dataset.

\section{Calibration on Gyro Fluid Solver (GFS) data} 
 Integrated modeling will be done using data generated from a linear solver such as the GFS, which is a more accurate linear eigensolver than the TGLF solver. Therefore, even though the linear quantities produced by the CGYRO are more accurate than the fast eigensolver GFS and we had already fit our neural network model to the more accurate CGYRO database, it is important that we test the model performance over GFS linear input data as well. To demonstrate the ability of the neural network model to predict 1D saturated potentials outside of the CGYRO dataset, the NN-SAT3 model is tested on the linear quantities produced by the GFS solver. The linear quantities generated from the GFS solver are over the same parameter scans that were used to generate the CGYRO linear quantities. In this section, we will discuss the model performance on GFS linear input data, and compare it with the results of the SAT3 model which is also tested on the same GFS data. Overall, there is good agreement between the linear quantities produced by CGYRO and GFS for low to medium $k_y$, whereas for high $k_y$ a mode transition is observed in the GFS data for the ITG cases. This difference in the linear data, as well as differences in accuracy and robustness between the SAT3 and NN-SAT3 models, lead to some differences in the output of the two models when tested on the GFS data, which we will analyze next.
 
\subsection{1D saturated potentials}\label{sec_gfs_flux}
\begin{figure}[!ht]
\centering
    \begin{subfigure}[b]{0.32\textwidth}
    \centering
    \includegraphics[width=\textwidth]{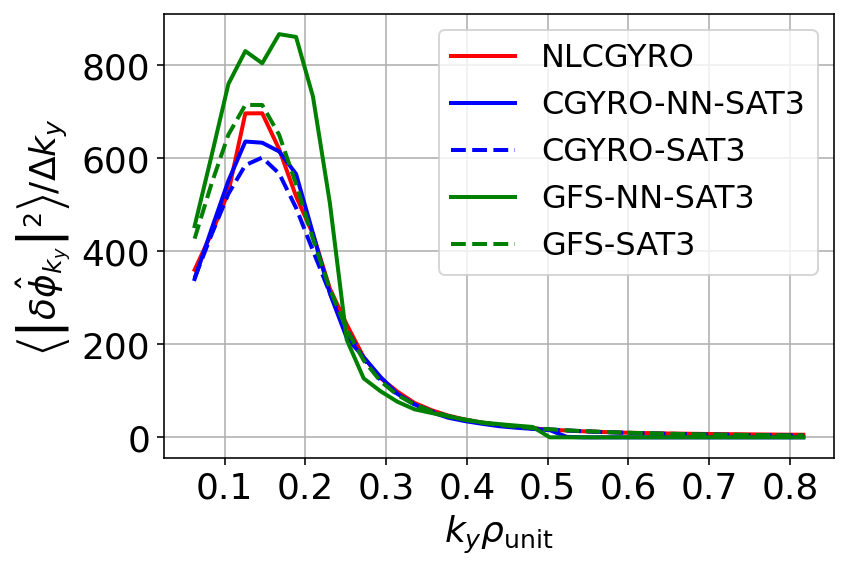}
    \caption{{\scriptsize $a/L_n = 3.0$, T}}
    \label{gfs1}
    \end{subfigure}
    \begin{subfigure}[b]{0.32\textwidth}
    \centering
    \includegraphics[width=\textwidth]{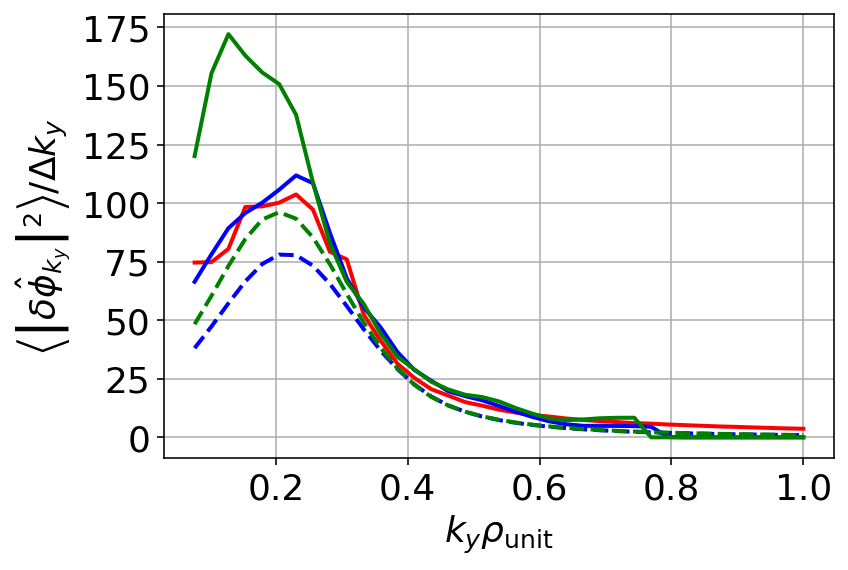}
    \caption{{\scriptsize $\hat{s} = 0.25$, D}}
    \label{gfs2}
    \end{subfigure}
    \begin{subfigure}[b]{0.32\textwidth}
    \centering
    \includegraphics[width=\textwidth]{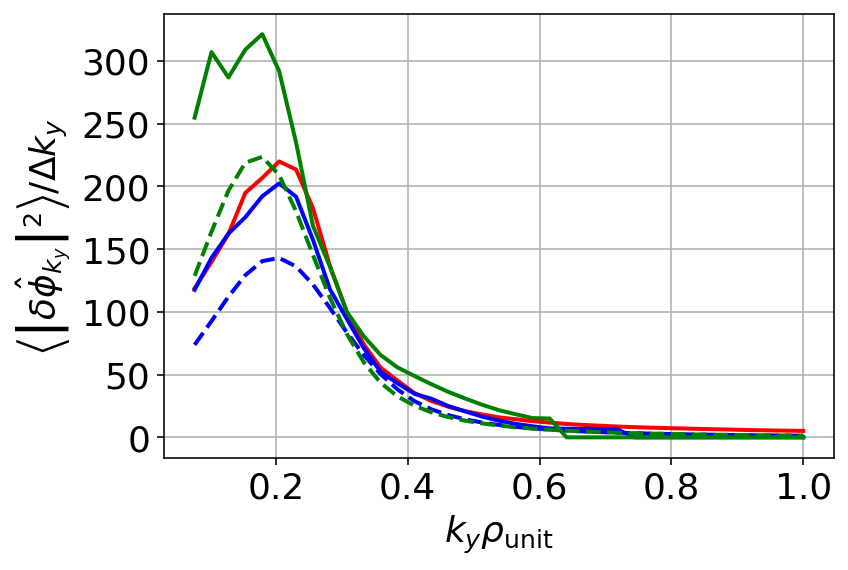}
    \caption{{\scriptsize $\hat{s} = 0.5$, D}}
    \label{gfs3}
    \end{subfigure}
    \begin{subfigure}[b]{0.32\textwidth}
    \centering
    \includegraphics[width=\textwidth]{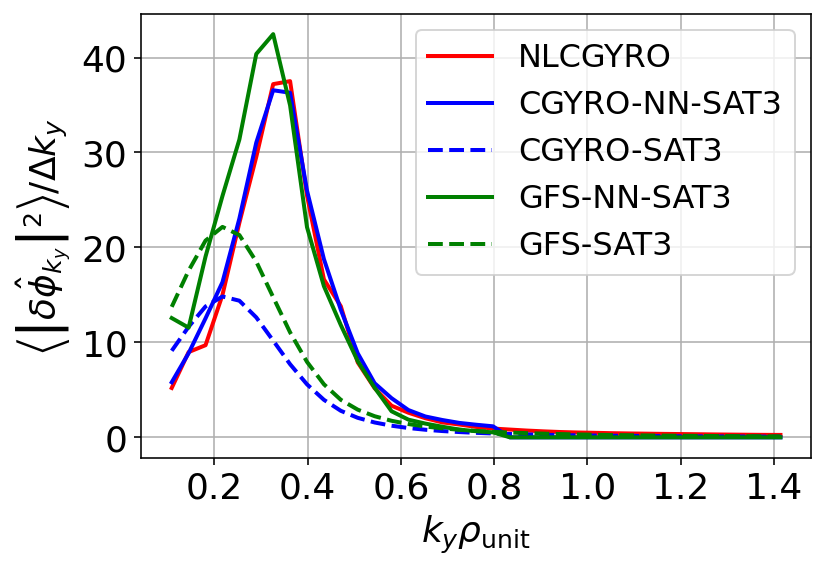}
    \caption{{\scriptsize $a/L_T = 1.5$, H}}
    \label{gfs4}
    \end{subfigure}
    \begin{subfigure}[b]{0.32\textwidth}
    \centering
    \includegraphics[width=\textwidth]{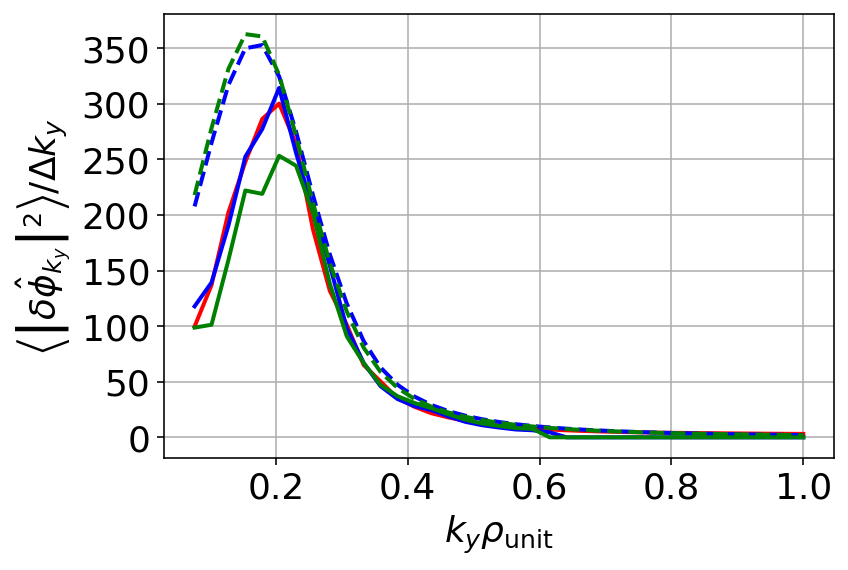}
    \caption{{\scriptsize $a/L_n = 2.0$, D}}
    \label{gfs5}
    \end{subfigure}
    \begin{subfigure}[b]{0.32\textwidth}
    \centering
    \includegraphics[width=\textwidth]{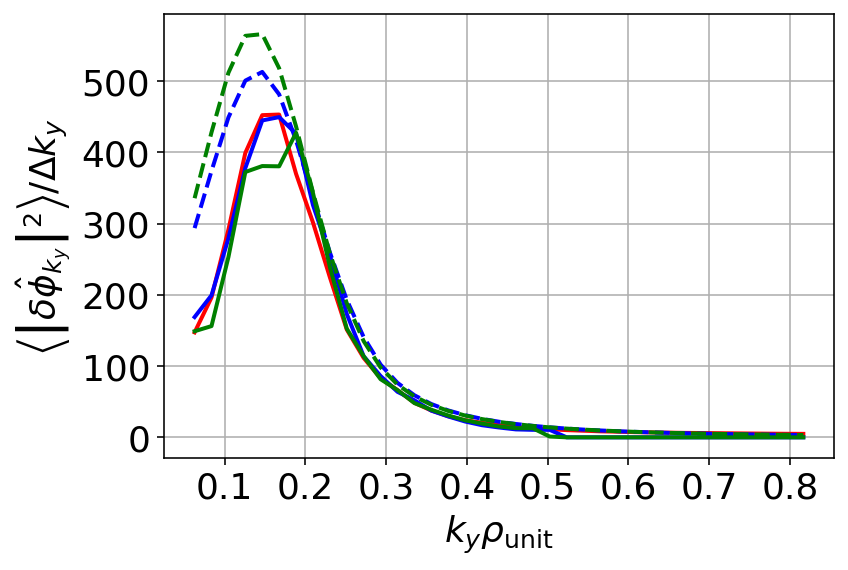}
    \caption{{\scriptsize $a/L_n = 2.0$, T}}
    \label{gfs6}
    \end{subfigure}
    \begin{subfigure}[b]{0.32\textwidth}
    \centering
    \includegraphics[width=\textwidth]{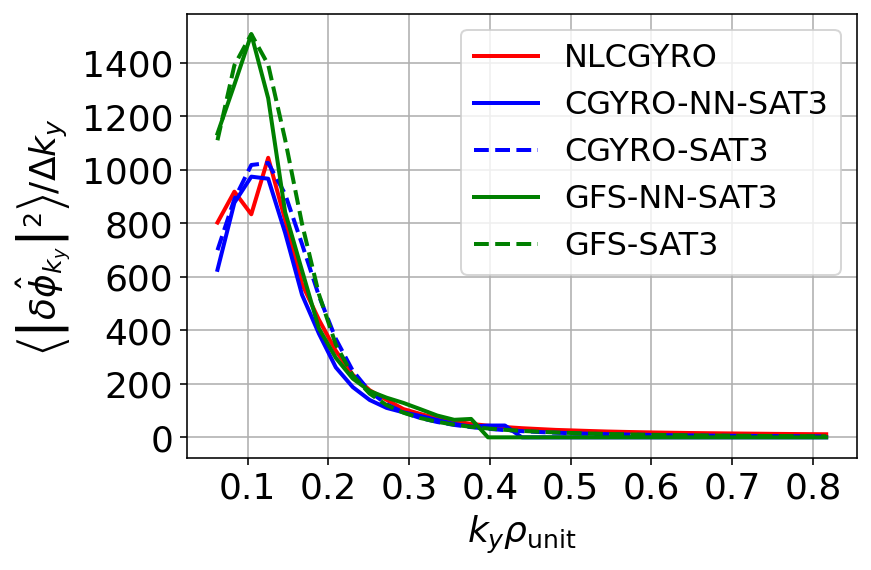}
    \caption{{\scriptsize $(a/c_s)\nu_{ee} = 0.01, a/L_n = 1.0$, T}}
    \label{gfs7}
    \end{subfigure}
    \begin{subfigure}[b]{0.32\textwidth}
    \centering
    \includegraphics[width=\textwidth]{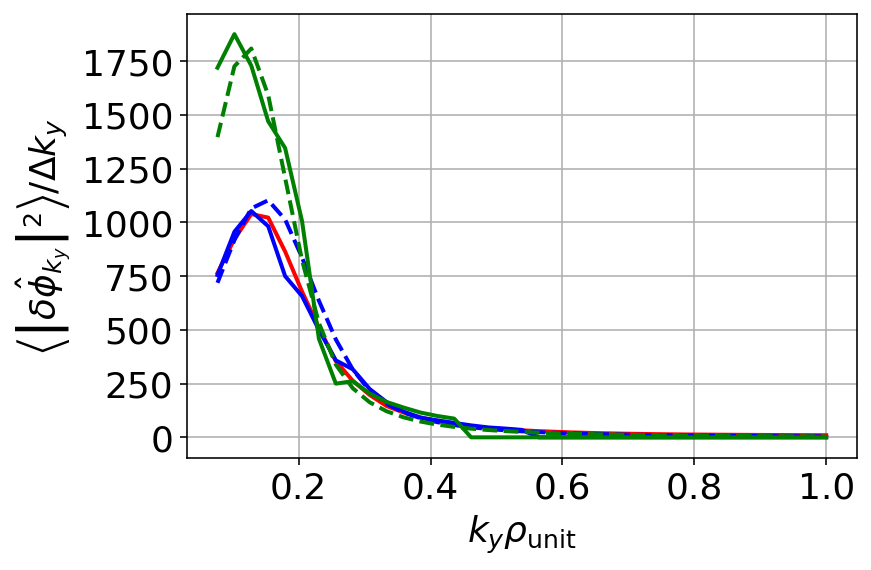}
    \caption{{\scriptsize $(a/c_s)\nu_{ee} = 0.01, a/L_n = 3.0$, D}}
    \label{gfs8}
    \end{subfigure}
    \begin{subfigure}[b]{0.32\textwidth}
    \centering
    \includegraphics[width=\textwidth]{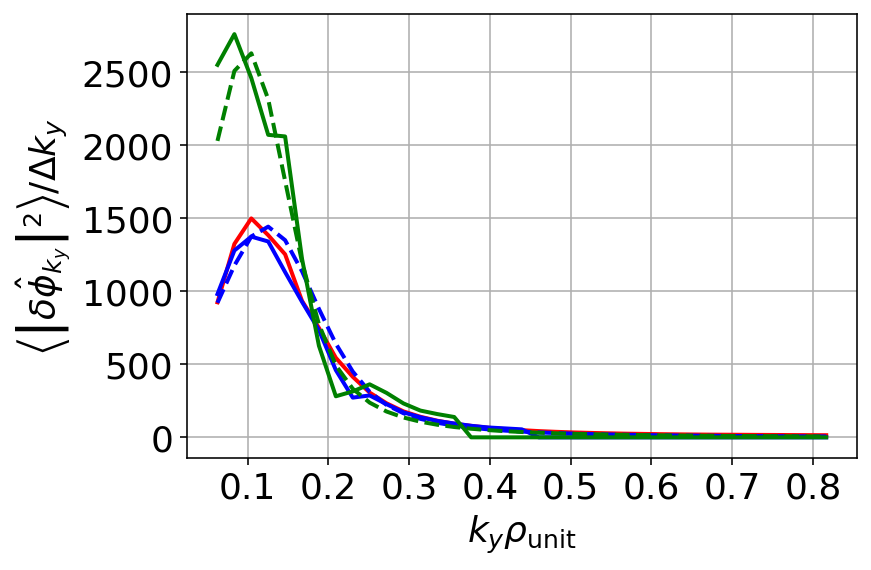}
    \caption{{\scriptsize $(a/c_s)\nu_{ee} = 0.01, a/L_n = 3.0$, T}}
    \label{gfs9}
    \end{subfigure}
    \captionsetup{justification=raggedright,singlelinecheck=false}
    \caption{1D saturated potentials for NL CGYRO (solid red), CGYRO-SAT3 (dashed blue), CGYRO-NN-SAT3 (solid blue), GFS-SAT3 (dashed green) and GFS-NN-SAT3 (solid green)  against binormal wavenumber $k_y$.}
    \label{gfs}
\end{figure}

The predictions of the 1D saturated potentials by GFS-SAT3 and GFS-NN-SAT3 for a few example cases are shown in figure \ref{gfs}. The 1D saturated potentials predicted by GFS-NN-SAT3 produce spikes for certain low $k_y$ for being near the zonal mode and also for some high $k_y$ because of mode transitions. The band pass filter, as mentioned in section \ref{arch}, suppresses these spikes.

First, we will discuss the performance of the two models for cases where GFS-NN-SAT3 over-predicts the 1D saturated potentials, as shown in the first row with figures \ref{gfs1}, \ref{gfs2} and \ref{gfs3}. 
This over-prediction can be attributed to the larger value of the linear inputs obtained from the GFS model such as frequency and linear weights for those cases. SAT3 does not use frequency and linear weights data as input and hence the 1D saturated potentials for the GFS-SAT3 are not affected by the changes in these inputs. 

Next, we analyze cases where GFS-NN-SAT3 performs better than GFS-SAT3, as shown in figures \ref{gfs4}, \ref{gfs5} and \ref{gfs6}. This is similar to the results we observed with CGYRO data and these results are consistent over the new GFS data, thus showing the improved performance for certain parameter scans regardless of the linear inputs, of the neural network model compared to a regression model like SAT3, which does not extrapolate accurately to new values of $\gamma_{max}$ and $k_{max}$.

Figures \ref{gfs7}, \ref{gfs8} and \ref{gfs9} show the plots for those cases in CGYRO which have high fluxes. Here, both GFS-NN-SAT3 and GFS-SAT3 over-predict the 1D saturated potentials. 
The similar performances of the two saturation models for inputs from a given linear solver (CGYRO or GFS), shows that the network performance in the high flux cases is an artifact of the linear solver type used for calculating the inputs rather than the saturation models.

\subsection{Fluxes}
\begin{figure}[!ht]
\centering
    \begin{subfigure}[b]{0.32\textwidth}
    \centering
    \includegraphics[width=\textwidth]{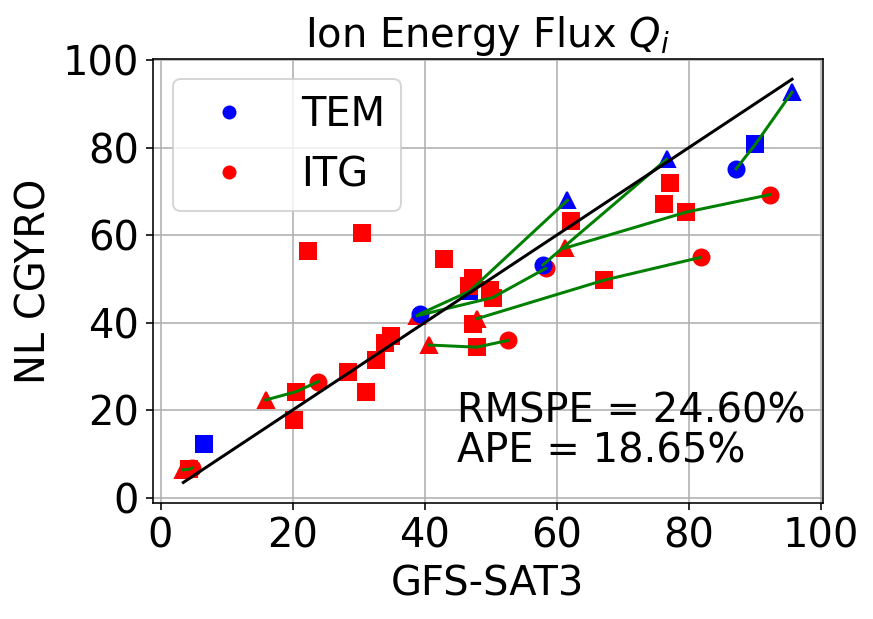}
    \caption{}
    \label{flux_qi_gfs_s3}
    \end{subfigure}
    \begin{subfigure}[b]{0.32\textwidth}
    \centering
    \includegraphics[width=\textwidth]{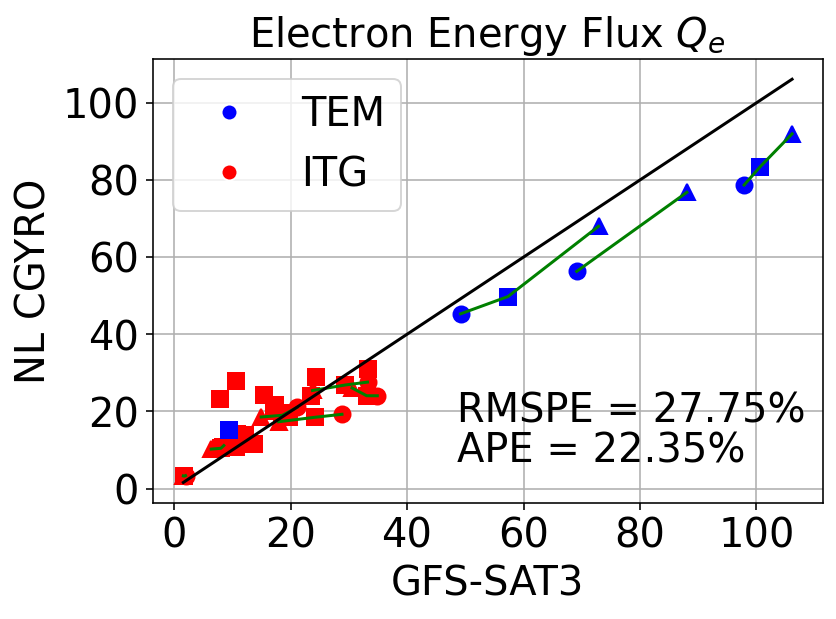}
    \caption{}
    \label{flux_qe_gfs_s3}
    \end{subfigure}
    \begin{subfigure}[b]{0.32\textwidth}
    \centering
    \includegraphics[width=\textwidth]{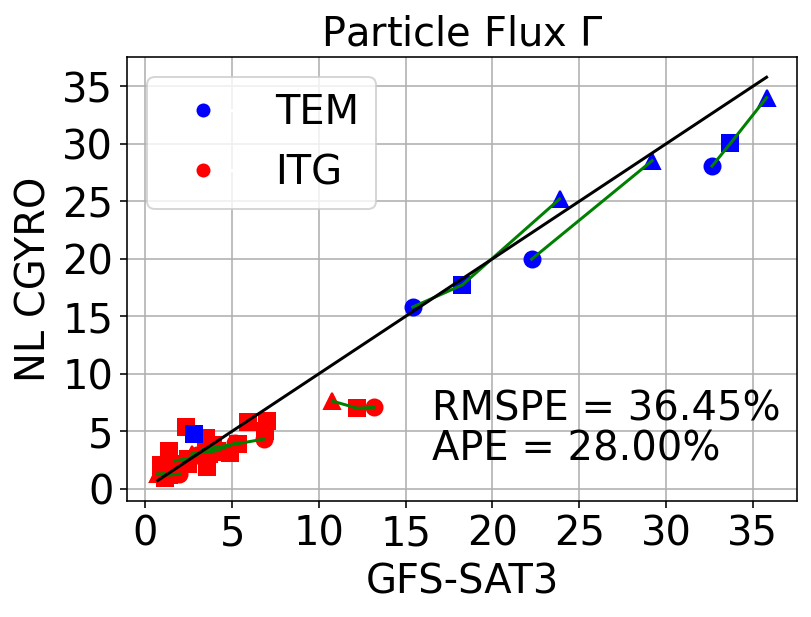}
    \caption{}
    \label{flux_qp_gfs_s3}
    \end{subfigure}
    \begin{subfigure}[b]{0.32\textwidth}
    \centering
    \includegraphics[width=\textwidth]{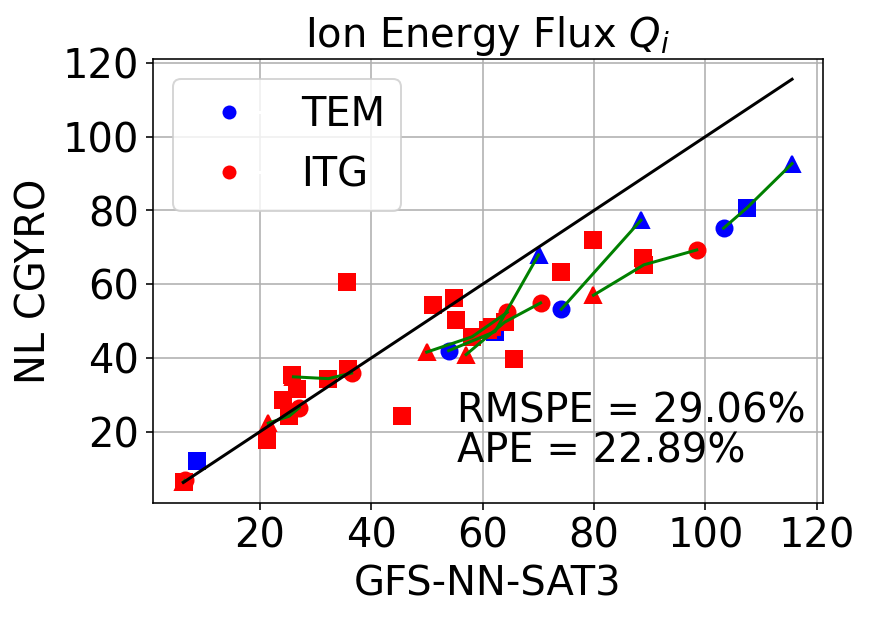}
    \caption{}
    \label{flux_qi_gfs_nns3}
    \end{subfigure}
    \begin{subfigure}[b]{0.32\textwidth}
    \centering
    \includegraphics[width=\textwidth]{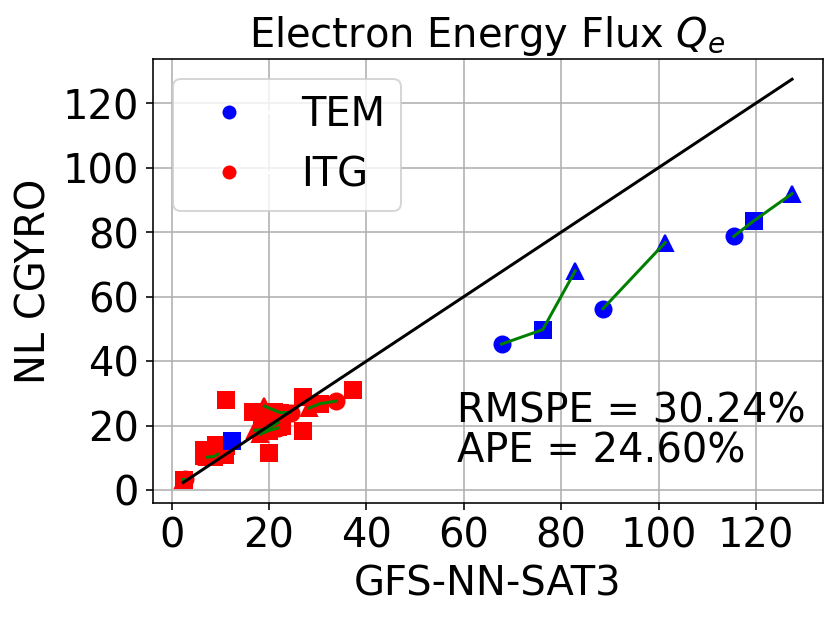}
    \caption{}
    \label{flux_qe_gfs_nns3}
    \end{subfigure}
    \begin{subfigure}[b]{0.32\textwidth}
    \centering
    \includegraphics[width=\textwidth]{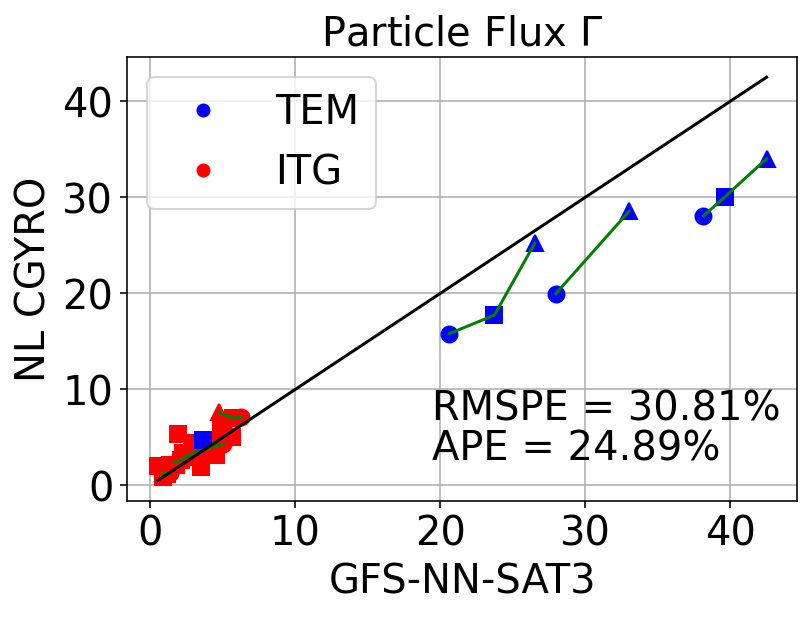}
    \caption{}
    \label{flux_qp_gfs_nns3}
    \end{subfigure}
    \captionsetup{justification=raggedright,singlelinecheck=false}
    \caption{Comparison of fluxes obtained from GFS-SAT3 and GFS-NN-SAT3 plotted against NL CGYRO data.}
    \label{q_gfs}
\end{figure}

Figure \ref{q_gfs} shows the total flux plots for $Q_i$, $Q_e$ and $\Gamma$, calculated using equation \ref{ql3} by substituting in the predicted 1D saturated potentials obtained using the linear inputs from GFS. The accuracy of $Q_i$, as shown in figures \ref{flux_qe_gfs_s3} and \ref{flux_qe_gfs_nns3}, is somewhat lesser for GFS-NN-SAT3 as compared to GFS-SAT3, where the APE has increased from 18.65\% to 22.89\% and the RMPSE has increased from 24.60\% to 29.06\%. For $Q_e$, as shown in figures \ref{flux_qe_gfs_s3} and \ref{flux_qe_gfs_nns3}, there is a slight increase in the APE from 22.35\% to 24.60\% and RMSPE from 27.75\% to 30.24\%, for the GFS-NN-SAT3 over GFS-SAT3. There is an improvement in the accuracy of $\Gamma$ for GFS-NN-SAT3 over GFS-SAT3, as shown in figures \ref{flux_qp_gfs_s3} and \ref{flux_qp_gfs_nns3}, where the APE has decreased from 28\% to 24.89\% and the RMSPE has decreased from 36.45\% to 30.81\%. Once again GFS-NN-SAT3 fits the points with density gradient $a/L_n = 2.0$ at the highest ITG particle fluxes, more accurately then GFS-SAT3, as was seen with the CGYRO data in figures \ref{flux_qp_s3} and \ref{flux_qp_nns3}.

\section{Conclusion}
The latest quasilinear saturation rule, SAT3 \cite{dudding2022new}, is an accurate model of the 1D saturated potential for an enhanced CGYRO database that encompasses the most common ITG and TEM turbulence regimes found in experiments and in particular captures relevant isotope scaling effects. It does not however represent a complete description of turbulent plasma transport and will perform less reliably in a parameter space far from its empirical regression. In this work, a generalized machine learning framework called the NN-SAT3, was formulated to model the magnitude of the 1D saturated potentials against binormal wavenumber from local linear simulations, using a neural network architecture trained on a database of high-resolution linea and nonlinear CGYRO simulations, which is the same as the one used in SAT3. NN-SAT3 is able to predict for linear inputs within the basis formed by the linear eigenmode input data it was trained on. Hence, the resulting neural network model generalizes well within ITG and TEM turbulence regimes. A machine learning fit that models the saturated potentials more accurately improves the calculation of the total energy and particle fluxes with quasilinear models. 

The fluxes obtained using this new CGYRO-NN-SAT3 model are not only more accurate, but also capture certain flux characteristics like anti-gyroBohm scaling for the TEM-dominated cases considered. CGYRO-NN-SAT3 also predicts more accurately near the point of critical gradient when plotted against temperature gradients. The overall accuracy of the ion energy and electron energy fluxes with respect to the NL CGYRO have improved significantly as compared to CGYRO-SAT3. There is some improvement in the particle flux accuracy over CGYRO-SAT3 but the density gradient cases $a/L_n = 2.0$ are captured far more accurately.

An additional database was generated from the GFS solver with the same input parameters that were used to generate the CGYRO database. Both the SAT3 and NN-SAT3 models were tested on this new database. We discussed plots of the 1D saturated potentials, with NN-SAT3 performing better than SAT3 in certain cases, showing consistent results as was observed with the CGYRO input database. We also showed cases where NN-SAT3 overpredicts the 1D saturated potentials and discussed the possible causes for the observation. We also discuss the total flux plots obtained using the GFS linear inputs, showing the ability of the models to generalize to a new linear solver. 

Current work is being done on integrated modeling studies using GFS-NN-SAT3 for analyzing its performance over previous saturation models, which will be published as part of a later work. For future work, the neural network model can be improved by training over more parameter scans in an expanded dataset obtained by adding more nonlinear gyrokinetic simulations to the current dataset. A case study is planned to test the NN-SAT3 model on new simulations and experimental data from JET, ITER and ASDEX Upgrade. This machine learning modeling framework can also be extended to include more challenging instabilities such as ETG and MTM leading to a more general model that can predict the turbulent transport effect of different modes.

\section*{Acknowledgment}
Research sponsored in part by the US Department of Energy under Contract DE-AC05-00OR22725 at Oak Ridge National Laboratory, managed by UT-Battelle, LLC for the Office of Science of the US Department of Energy.


\bibliographystyle{cas-model2-names}

\bibliography{cas-refs}

@PREAMBLE{
 "\providecommand{\noopsort}[1]{}" 
 # "\providecommand{\singleletter}[1]{#1}%" 
}

@article{staebler2021verification,
  title={Verification of a quasi-linear model for gyrokinetic turbulent transport},
  author={Staebler, Gary M and Belli, EA and Candy, J and Kinsey, JE and Dudding, H and Patel, B},
  journal={Nuclear Fusion},
  volume={61},
  number={11},
  pages={116007},
  year={2021},
  publisher={IOP Publishing}
}

@article{candy2016high,
  title={A high-accuracy Eulerian gyrokinetic solver for collisional plasmas},
  author={Candy, Jeff and Belli, Emily A and Bravenec, RV},
  journal={Journal of Computational Physics},
  volume={324},
  pages={73--93},
  year={2016},
  publisher={Elsevier}
}

@article{belli2019reversal,
  title={Reversal of turbulent gyroBohm isotope scaling due to nonadiabatic electron drive},
  author={Belli, EA and Candy, J and Waltz, RE},
  journal={Physics of Plasmas},
  volume={26},
  number={8},
  year={2019},
  publisher={AIP Publishing}
}

@article{garcia2022modelling,
  title={Modelling and theoretical understanding of the isotope effect from JET experiments in view of reliable predictions for deuterium-tritium plasmas},
  author={Garcia, J and Casson, FJ and Navarro, A Ba{\~n}{\'o}n and Bonanomi, N and Citrin, J and King, D and Mantica, P and Mariani, A and Marin, M and Mazzi, S and others},
  journal={Plasma Physics and Controlled Fusion},
  volume={64},
  number={5},
  pages={054001},
  year={2022},
  publisher={IOP Publishing}
}

@article{rhodes2011mode,
  title={L-mode validation studies of gyrokinetic turbulence simulations via multiscale and multifield turbulence measurements on the DIII-D tokamak},
  author={Rhodes, TL and Holland, C and Smith, SP and White, AE and Burrell, KH and Candy, J and DeBoo, JC and Doyle, EJ and Hillesheim, JC and Kinsey, JE and others},
  journal={Nuclear Fusion},
  volume={51},
  number={6},
  pages={063022},
  year={2011},
  publisher={IOP Publishing}
}

@article{holland2011advances,
  title={Advances in validating gyrokinetic turbulence models against L-and H-mode plasmas},
  author={Holland, C and Schmitz, L and Rhodes, TL and Peebles, WA and Hillesheim, JC and Wang, G and Zeng, L and Doyle, EJ and Smith, SP and Prater, R and others},
  journal={Physics of Plasmas},
  volume={18},
  number={5},
  year={2011},
  publisher={AIP Publishing}
}

@article{told2013characterizing,
  title={Characterizing turbulent transport in ASDEX Upgrade L-mode plasmas via nonlinear gyrokinetic simulations},
  author={Told, D and Jenko, F and G{\"o}rler, T and Casson, FJ and Fable, E and ASDEX Upgrade Team and others},
  journal={Physics of Plasmas},
  volume={20},
  number={12},
  year={2013},
  publisher={AIP Publishing}
}

@article{citrin2014ion,
  title={Ion temperature profile stiffness: non-linear gyrokinetic simulations and comparison with experiment},
  author={Citrin, J and Jenko, F and Mantica, P and Told, D and Bourdelle, C and Dumont, R and Garcia, J and Haverkort, JW and Hogeweij, GMD and Johnson, Thomas and others},
  journal={Nuclear Fusion},
  volume={54},
  number={2},
  pages={023008},
  year={2014},
  publisher={IOP Publishing}
}

@article{bonanomi2015trapped,
  title={Trapped electron mode driven electron heat transport in JET: experimental investigation and gyro-kinetic theory validation},
  author={Bonanomi, N and Mantica, P and Szepesi, G and Hawkes, N and Lerche, E and Migliano, P and Peeters, A and Sozzi, C and Tsalas, M and Van Eester, D and others},
  journal={Nuclear Fusion},
  volume={55},
  number={11},
  pages={113016},
  year={2015},
  publisher={IOP Publishing}
}

@article{creely2017validation,
  title={Validation of nonlinear gyrokinetic simulations of L-and I-mode plasmas on Alcator C-Mod},
  author={Creely, AJ and Howard, NT and Rodriguez-Fernandez, P and Cao, N and Hubbard, AE and Hughes, JW and Rice, JE and White, AE and Candy, J and Staebler, GM and others},
  journal={Physics of Plasmas},
  volume={24},
  number={5},
  year={2017},
  publisher={AIP Publishing}
}

@article{freethy2018validation,
  title={Validation of gyrokinetic simulations with measurements of electron temperature fluctuations and density-temperature phase angles on ASDEX Upgrade},
  author={Freethy, SJ and G{\"o}rler, T and Creely, AJ and Conway, GD and Denk, SS and Happel, T and Koenen, C and Hennequin, Pascale and White, AE and ASDEX Upgrade Team and others},
  journal={Physics of Plasmas},
  volume={25},
  number={5},
  year={2018},
  publisher={AIP Publishing}
}

@article{citrin2022integrated,
  title={Integrated modelling and multiscale gyrokinetic validation study of ETG turbulence in a JET hybrid H-mode scenario},
  author={Citrin, J and Maeyama, S and Angioni, C and Bonanomi, N and Bourdelle, C and Casson, FJ and Fable, E and G{\"o}rler, T and Mantica, P and Mariani, A and others},
  journal={Nuclear Fusion},
  volume={62},
  number={8},
  pages={086025},
  year={2022},
  publisher={IOP Publishing}
}

@article{staebler2005gyro,
  title={Gyro-Landau fluid equations for trapped and passing particles},
  author={Staebler, GM and Kinsey, JE and Waltz, RE},
  journal={Physics of Plasmas},
  volume={12},
  number={10},
  year={2005},
  publisher={AIP Publishing}
}

@article{bourdelle2015core,
  title={Core turbulent transport in tokamak plasmas: bridging theory and experiment with QuaLiKiz},
  author={Bourdelle, C and Citrin, J and Baiocchi, B and Casati, A and Cottier, P and Garbet, X and Imbeaux, F and Contributors, JET},
  journal={Plasma Physics and Controlled Fusion},
  volume={58},
  number={1},
  pages={014036},
  year={2015},
  publisher={IOP Publishing}
}

@article{staebler2016role,
  title={The role of zonal flows in the saturation of multi-scale gyrokinetic turbulence},
  author={Staebler, Gary M and Candy, John and Howard, Nathan T and Holland, Christopher},
  journal={Physics of Plasmas},
  volume={23},
  number={6},
  year={2016},
  publisher={AIP Publishing}
}

@article{casati2009validating,
  title={Validating a quasi-linear transport model versus nonlinear simulations},
  author={Casati, A and Bourdelle, C and Garbet, X and Imbeaux, Fr{\'e}d{\'e}ric and Candy, J and Clairet, F and Dif-Pradalier, Guilhem and Falchetto, G and Gerbaud, T and Grandgirard, Virginie and others},
  journal={Nuclear Fusion},
  volume={49},
  number={8},
  pages={085012},
  year={2009},
  publisher={IOP Publishing}
}

@article{waltz1997gyro,
  title={A gyro-Landau-fluid transport model},
  author={Waltz, RE and Staebler, GM and Dorland, W and Hammett, GW and Kotschenreuther, Mike and Konings, JA},
  journal={Physics of Plasmas},
  volume={4},
  number={7},
  pages={2482--2496},
  year={1997},
  publisher={American Institute of Physics}
}

@article{candy2009unified,
  title={A unified method for operator evaluation in local Grad--Shafranov plasma equilibria},
  author={Candy, J},
  journal={Plasma Physics and Controlled Fusion},
  volume={51},
  number={10},
  pages={105009},
  year={2009},
  publisher={IOP Publishing}
}

@techreport{lee1984nonlinear,
  title={Nonlinear evolution of drift instabilities},
  author={Lee, WW and Krommes, John A and Oberman, Carl R and Smith, Ralph A},
  year={1984},
  institution={Princeton Plasma Physics Lab.(PPPL), Princeton, NJ (United States)}
}

@article{dudding2022new,
  title={A new quasilinear saturation rule for tokamak turbulence with application to the isotope scaling of transport},
  author={Dudding, Harry George and Casson, FJ and Dickinson, David and Patel, BS and Roach, CM and Belli, EA and Staebler, GM},
  journal={Nuclear Fusion},
  volume={62},
  number={9},
  pages={096005},
  year={2022},
  publisher={IOP Publishing}
}

@article{staebler2007theory,
  title={A theory-based transport model with comprehensive physics},
  author={Staebler, GM and Kinsey, JE and Waltz, RE},
  journal={Physics of Plasmas},
  volume={14},
  number={5},
  year={2007},
  publisher={AIP Publishing}
}

@article{kinsey2008first,
  title={The first transport code simulations using the trapped gyro-Landau-fluid model},
  author={Kinsey, JE and Staebler, GM and Waltz, RE},
  journal={Physics of Plasmas},
  volume={15},
  number={5},
  year={2008},
  publisher={AIP Publishing}
}

@article{fransson2023fast,
  title={A fast neural network surrogate model for the eigenvalues of QuaLiKiz},
  author={Fransson, E and Gillgren, A and Ho, A and Borsander, J and Lindberg, O and Rieck, W and {\AA}qvist, M and Strand, P},
  journal={Physics of Plasmas},
  volume={30},
  number={12},
  year={2023},
  publisher={AIP Publishing}
}

@article{van2020fast,
  title={Fast modeling of turbulent transport in fusion plasmas using neural networks},
  author={van de Plassche, Karel Lucas and Citrin, Jonathan and Bourdelle, Clarisse and Camenen, Yann and Casson, Francis J and Dagnelie, Victor I and Felici, Federico and Ho, Aaron and Van Mulders, Simon and JET, Contributors},
  journal={Physics of Plasmas},
  volume={27},
  number={2},
  year={2020},
  publisher={AIP Publishing}
}

@article{citrin2015real,
  title={Real-time capable first principle based modelling of tokamak turbulent transport},
  author={Citrin, Jonathan and Breton, Sarah and Felici, Federico and Imbeaux, Frederic and Aniel, T and Artaud, JF and Baiocchi, B and Bourdelle, C and Camenen, Y and Garcia, J},
  journal={Nuclear Fusion},
  volume={55},
  number={9},
  pages={092001},
  year={2015},
  publisher={IOP Publishing}
}

@article{cao2025tglf,
  title={TGLF-SINN: Deep Learning Surrogate Model for Accelerating Turbulent Transport Modeling in Fusion},
  author={Cao, Yadi and Zhang, Futian and Liu, Wesley and Neiser, Tom and Meneghini, Orso and Fuller, Lawson and Smith, Sterling and Nazikian, Raffi and Sammuli, Brian and Yu, Rose},
  journal={arXiv preprint arXiv:2509.07024},
  year={2025}
}

@article{meneghini2017self,
  title={Self-consistent core-pedestal transport simulations with neural network accelerated models},
  author={Meneghini, Orso and Smith, Sterling P and Snyder, Philip B and Staebler, Gary M and Candy, Jeffrey and Belli, E and Lao, L and Kostuk, Mark and Luce, T and Luda, Teobaldo and others},
  journal={Nuclear Fusion},
  volume={57},
  number={8},
  pages={086034},
  year={2017},
  publisher={IOP Publishing}
}

@inproceedings{neiser2022database,
  title={Database generation for validation of TGLF and retraining of neural network accelerated TGLF-NN},
  author={Neiser, Tom and Meneghini, Orso and Smith, Sterling and McClenaghan, Joseph and Orozco, David and Hall, Joseph and Staebler, Gary and Belli, Emily and Candy, Jeff},
  booktitle={APS Division of Plasma Physics Meeting Abstracts},
  volume={2022},
  pages={GP11--010},
  year={2022}
}

@misc{citekey,
  author = {Harry Dudding},
  title = {SAT3 Dataset},
  howpublished = "\url{https://zenodo.org/records/14505760}",
  year = {2024}
}

@article{kinsey2010trapped,
  title={Trapped gyro-Landau-fluid transport modeling of DIII-D hybrid discharges},
  author={Kinsey, JE and Staebler, GM and Petty, CC},
  journal={Physics of Plasmas},
  volume={17},
  number={12},
  year={2010},
  publisher={AIP Publishing}
}

@article{staebler2023flexible,
  title={A flexible gyro-fluid system of equations},
  author={Staebler, Gary M and Belli, EA and Candy, Jeff},
  journal={Physics of Plasmas},
  volume={30},
  number={10},
  year={2023},
  publisher={AIP Publishing}
}

@article{dudding2026integrated,
  title={Integrated modelling simulations of Ohmic and L-mode JET discharges in H, D and T using JETTO-TGLF},
  author={Dudding, Harry George and Casson, Francis J and Dickinson, David and Roach, Colin M and Patel, Bhavin S and Bache, Tom W and Delabie, Ephrem and Maggi, Costanza F and Rodriguez-Fernandez, Pablo and Nave, Maria Filomena Ferreira},
  journal={Plasma Physics and Controlled Fusion},
  year={2026}
}



\end{document}